\documentclass{aastex701}

\usepackage{booktabs}
\usepackage{tabularx}
\usepackage{siunitx}

\shorttitle{DIPS-OjOS I}
\shortauthors{Díez Alonso et al.}

\submitjournal{PASP}

\renewcommand{\today}{May 29, 2026}
\begin{document}

\title{DIPS-OjOS I: Project Design and Science Validation}

\author[0000-0002-5826-9892]{E. Díez Alonso}
\affiliation{Instituto Universitario de Ciencias y Tecnologías Espaciales de Asturias (ICTEA), C. Independencia 13, 33004 Oviedo, Spain}
\affiliation{Departamento de Matemáticas, Facultad de Ciencias, Universidad de Oviedo, 33007 Oviedo, Spain}
\affiliation{Sociedad Astronómica Asturiana Omega, C. Alegría 45, 33209 Gijón, Spain}
\email[show]{diezenrique@uniovi.es}
\correspondingauthor{E. Díez Alonso}

\author[0000-0001-5903-9899]{S. Iglesias Alvarez}
\affiliation{Instituto Universitario de Ciencias y Tecnologías Espaciales de Asturias (ICTEA), C. Independencia 13, 33004 Oviedo, Spain}
\email{iglesiassantiago@uniovi.es}

\author{J. Menéndez Blanco}
\affiliation{Sociedad Astronómica Asturiana Omega, C. Alegría 45, 33209 Gijón, Spain}
\email{juanmb@gmail.com}

\author{J. Collada Bárcena}
\affiliation{Sociedad Astronómica Asturiana Omega, C. Alegría 45, 33209 Gijón, Spain}
\email{jncollada@gmail.com}

\author{A. Coya Lozano}
\affiliation{Sociedad Astronómica Asturiana Omega, C. Alegría 45, 33209 Gijón, Spain}
\email{alfonsocoya@gmail.com}

\author[0009-0007-2419-2405]{J. M. Fernández Díaz}
\affiliation{Instituto Universitario de Ciencias y Tecnologías Espaciales de Asturias (ICTEA), C. Independencia 13, 33004 Oviedo, Spain}
\email{fernandezjulia@uniovi.es}

\author{D. García}
\affiliation{Sociedad Astronómica Asturiana Omega, C. Alegría 45, 33209 Gijón, Spain}
\email{danigarcia77@gmail.com}

\author{F. García de la Cuesta}
\affiliation{Sociedad Astronómica Asturiana Omega, C. Alegría 45, 33209 Gijón, Spain}
\email{faustino.garcia@gmail.com}

\author{N. Gómez Hernández}
\affiliation{Sociedad Astronómica Asturiana Omega, C. Alegría 45, 33209 Gijón, Spain}
\email{nabi_astur@hotmail.es}

\author{Á. J. González Ordóñez}
\affiliation{Sociedad Astronómica Asturiana Omega, C. Alegría 45, 33209 Gijón, Spain}
\email{jagonzalezor@gmail.com}

\author{R. Hevia Díaz}
\affiliation{Instituto Universitario de Ciencias y Tecnologías Espaciales de Asturias (ICTEA), C. Independencia 13, 33004 Oviedo, Spain}
\email{ramonheviadiaz@gmail.com}

\author{R. Pardo Silva}
\affiliation{Sociedad Astronómica Asturiana Omega, C. Alegría 45, 33209 Gijón, Spain}
\email{raulgijon@gmail.com}

\author{J. Polanco}
\affiliation{Sociedad Astronómica Asturiana Omega, C. Alegría 45, 33209 Gijón, Spain}
\email{javierpolancos@hotmail.es}

\author{A. Quijada Espina}
\affiliation{Sociedad Astronómica Asturiana Omega, C. Alegría 45, 33209 Gijón, Spain}
\email{toquies2002@yahoo.es}

\author[0000-0002-2718-2022]{J. Rodríguez Rodríguez}
\affiliation{Instituto Universitario de Ciencias y Tecnologías Espaciales de Asturias (ICTEA), C. Independencia 13, 33004 Oviedo, Spain}
\affiliation{Departamento de Matemáticas, Facultad de Ciencias, Universidad de Oviedo, 33007 Oviedo, Spain}
\email{rodriguezrjavier@uniovi.es}

\author{N. Villanueva Fernández}
\affiliation{Sociedad Astronómica Asturiana Omega, C. Alegría 45, 33209 Gijón, Spain}
\email{nvillanuevafernandez9@gmail.com}

\author[0000-0002-9660-7944]{F. J. de Cos Juez}
\affiliation{Instituto Universitario de Ciencias y Tecnologías Espaciales de Asturias (ICTEA), C. Independencia 13, 33004 Oviedo, Spain}
\email{fjcos@uniovi.es}

\begin{abstract}
Automated pipelines and machine-learning algorithms are efficient at detecting standard planetary transits, but can miss subtle, non-periodic morphological anomalies. We present the pilot phase of DIPS-OjOS (Detecting Irregular Photometric Signals -- OjímetrO Survey), a professional-amateur collaboration designed to develop and validate a scalable workflow for identifying irregular photometric signals in exoplanet transit light curves through trained visual inspection. The project targets starspot-crossing events, stellar flares, circumplanetary structures, exomoon-like signatures, and transit-duration or timing anomalies overlooked by automated searches.

We built the infrastructure required for a controlled visual-inspection experiment: simulated control light curves, preparation and detrending of real short-cadence \textit{Kepler} data, a transit-modeling pipeline that computes an optimized average transit profile for each light curve, and a web-based classification platform. The real sample was deliberately restricted to 51 well-characterized \textit{Kepler} systems, providing a conservative testbed rather than a discovery-optimized survey. Collaborators inspected 917 transit events: 555 real transits and 362 simulated controls. The simulated sample quantified observer performance and calibrated a weighted consensus scheme; collaborators recovered injected signals with an average efficiency of 87.6\%, while the null-case false-positive rate was 15.3\%.

After expert validation, we retained 88 viable candidate anomalies from the real sample, including 26 starspot-crossing event candidates across eight targets, seven stellar flares, 31 systematic transit-duration deviations motivating refined orbital parameters in four systems, and a mutual transit event in Kepler-89. Recovering candidate signals in this well-studied sample validates the sensitivity of DIPS-OjOS and establishes the basis for future applications to \textit{Kepler}, TESS, PLATO, and ground-based datasets.
\end{abstract}


\section{Introduction}

Our understanding of planetary systems has grown exponentially in the present century, driven by the success of search programs based on radial velocity (HIRES, \citealp{1994SPIE.2198..362V}; HARPS, \citealp{2003Msngr.114...20M}; CARMENES, \citealp{2018SPIE10702E..0WQ}, SPRESSO \citealp{2021A&A...645A..96P}) and photometry, both ground-based (OGLE, \citealp{1992AcA....42..253U}; SuperWASP, \citealp{2006PASP..118.1407P}; MEarth, \citealp{2008PASP..120..317N}) and space-borne (\textit{Kepler}, \citealp{2010Sci...327..977B}; TESS, \citealp{2014SPIE.9143E..20R}).

Astrophysically valuable anomalies can be embedded within planetary transits. These anomalies largely originate from stellar activity, particularly the presence of active regions and starspots. Starspot-crossing events (SSCEs) have recently been the subject of systematic searches in light curves \citep{2025A&A...704A.102H}. Their accurate characterization is essential, as SSCEs can be confused with rings or satellites \citep{2007A&A...476.1347P}, introduce noise in the measure of transit timing variations (TTVs, \citealp{2015ApJ...800..142M}), and significantly alter the planetary parameters derived from transit fits \citep{2009A&A...505.1277C,2013A&A...556A..19O}.

Furthermore, the detailed analysis of these anomalies provides a tool to map the surface distribution of active regions \citep{2003ApJ...585L.147S, 2012A&A...547A..37B} and derive crucial dynamical parameters, such as the direction of planetary orbits or stellar obliquity \citep{2011ApJS..197...14D,2015ApJ...800..142M}.

SSCEs are closely linked to another manifestation of magnetic activity: stellar flares, also target of systematic searches \citep{2016ApJ...829...23D, 2017ApJS..232...26V,2019ApJS..241...29Y}.
The joint analysis of both phenomena allows us to establish correlations between the spot surface area and the energy released during these events \citep{2021ApJ...922L..23A}. Moreover, a serendipitous planetary occultation of a flaring region offers a unique opportunity to spatially constrain its location \citep{2025MNRAS.538.2937A}, which, together with flare energy and frequency, is crucial for understanding the impact of flare-induced erosion on the evolution of planetary atmospheres \citep{2017A&A...608A..75C}.

The impact of this magnetic activity is equally critical for exoplanet searches using the radial velocity method, where signals induced by spot rotation can easily mimic the presence of exoplanets \citep{2001A&A...379..279Q,2007A&A...473..983D}. Crucially, characterizing these features is essential to prevent biased estimations of planetary properties and to mitigate the impact of stellar activity in radial velocity follow-up surveys.

Signatures originated by the presence of rings \citep{2004ApJ...616.1193B,2012AJ....143...72M} or circumplanetary dust \citep{2012ApJ...752....1R, 2015ApJ...800..126K} can also be detected in photometric transits. Furthermore, visual inspection can reveal transit duration variations (TDVs) caused by dynamical interactions \citep{2009MNRAS.392..181K, 2013ApJ...777....3N}, mutual phenomena between overlapping planets \citep{2012ApJ...759L..36H}, or subtle morphological anomalies that may hint at the existence of previously undetected bodies, such as additional planets or exomoons \citep{2012ApJ...750..115K,2018SciA....4.1784T}. Additionally, the detection of systematic anomalies can suggest incorrect values in the tabulated parameters for the exoplanets or for the stellar limb darkening.

To handle the sheer volume of data generated by photometric surveys, automated pipelines such as the Box-Least Squares algorithm (BLS,  \citealp{2002A&A...391..369K}) and the more recent Transit Least Squares (TLS, \citealp{2019A&A...623A..39H}) have been implemented to detect and characterize transiting exoplanetary candidates. Furthermore, automated vetting algorithms (such as the Kepler Robovetter; \citealp{2018ApJS..235...38T}) and artificial intelligence architectures \citep{2018AJ....155...94S, 2023arXiv231207161I, 2024arXiv240213673I} have been deployed to process these light curves efficiently. Nevertheless, these automated methods struggle to detect and classify subtle signals present in the light curves.

While significantly more time-intensive, visual inspection by trained observers has proven capable of identifying these elusive signals. Previous successful citizen science initiatives, such as Planet Hunters \citep{2012MNRAS.419.2900F, 2016MNRAS.457.3988B}, TESS Hunters \citep{2020MNRAS.494..750E}, Planet Patrol \citep{2022PASP..134d4401K} and K2-OjOS \citep{2022MNRAS.509.1075C}, have primarily focused on the discovery of new exoplanets through by-eye examination of light curves.

In this paper, we present the pilot phase of the DIPS-OjOS project (Detecting Irregular Photometric Signals -- OjímetrO Survey), a professional-amateur (pro-am) initiative led by researchers from the Instituto Universitario de Ciencias y Tecnologías Espaciales de Asturias (ICTEA) in partnership with members of the K2-OjOS\footnote{\url{https://sites.google.com/view/k2ojos/}} project. The goal of this first phase is not to produce an exhaustive catalog of transit anomalies, but to validate the observational, technical, and human-classification workflow required for a larger-scale search.

To this end, we developed the infrastructure required to carry out a controlled visual-inspection experiment. This includes the generation of simulated control light curves, the preparation and detrending of real short-cadence \textit{Kepler} data, a transit-modeling pipeline that computes an optimized average transit profile for each light curve, and a web-based platform through which trained collaborators inspected and classified individual events. The real sample was deliberately restricted to 51 well-characterized \textit{Kepler} systems, providing a conservative testbed for the methodology rather than a discovery-optimized survey. In total, collaborators inspected 917 individual transit events, combining the simulated control sample, used to quantify recovery efficiency and false-positive rates, with the real \textit{Kepler} sample. The candidate signals recovered from this well-studied dataset therefore provide a first validation of the sensitivity of the workflow and establish the methodological and operational framework for scaling DIPS-OjOS to larger datasets from the \textit{Kepler}, TESS, and PLATO \citep{2014ExA....38..249R} missions, as well as ground-based surveys.

\section{Methods}

\subsection{Real Sample}

For this pilot phase, we selected a manageable sample of 51 targets, including both confirmed and candidate systems, observed during the original \textit{Kepler} mission. To ensure sufficient temporal resolution the selection was restricted to targets observed in short cadence. Furthermore, to analyze signals with high signal-to-noise ratios, we required a transit depth $\delta > 1000$ ppm and a host star magnitude $K_{\rm p}<14$. We also imposed an orbital period constraint of $P_{\rm orb}>20$ d. This threshold also includes the regime where dynamical conditions allow for the presence of orbiting satellites, thus avoiding the exclusion of systems that could potentially harbor exomoons. As an exception, we included the transits of Kepler-411 b and c due to the high activity of their host star and because their data were already contained within the light curves of planet d.

After filtering the sample of planets, we obtained the following planetary parameters to build the theoretical transit models: orbit period ($P_{\rm orb}$), transit epoch, transit duration, planet radius ($R_{\rm p}$), orbital semi-major axis ($a$), inclination ($i$), eccentricity ($e$) and longitude of periastron ($\omega$). Likewise, the following stellar parameters were downloaded: star radius ($R_{\star}$), effective temperature ($T_{\rm eff}$), surface gravity ($\log g$) and metallicity ($[Fe/H]$). For those confirmed planets with \textit{Kepler} ID, we downloaded all the parameters from the NASA Exoplanet Archive (NEA, \citealp{2013PASP..125..989A}) via \texttt{astroquery} \citep{2019AJ....157...98G}, while for Kepler Objects of Interests (KOIs) we obtained the planet and stellar parameters from \citep{2018ApJS..235...38T} and \citep{2019AJ....158..138S} respectively. Finally, we derived the quadratic limb-darkening coefficients $u_{1}$ and $u_{2}$ by interpolating the stellar parameters from the tables provided by \citet{2017A&A...600A..30C}.

Next, we used the \texttt{lightkurve} package \citep{2018ascl.soft12013L} to download all available short-cadence photometry for each target. We partitioned the data into separate segments whenever we identified gaps exceeding five days. This process resulted in a total of 140 segments, comprising 555 individual transits. Finally, we used the \texttt{wotan} package \citep{2019AJ....158..143H} to remove long-term trends and stellar variability from each light curve, performing this detrending process individually for each segment, carefully selecting the window\_length parameter to preserve the transit morphology and avoid signal distortion. In Table~\ref{tab:real_exoplanets} we present the 51 targets along with their parameters.

\begin{table}[p]
\centering
\small
\setlength{\tabcolsep}{2pt}
\caption{Summary of the \textit{Kepler} planetary sample inspected in this work.} 
\label{tab:real_exoplanets}
\begin{tabular}{lcccccc}
\toprule
\toprule
Exoplanet &
$R_{\rm p}/R_{\star}$ &
$a/R_{\star}$ &
$i$ ($^\circ$) &
$P$ (d) &
\# Seg. &
\# Tr. \\
\midrule
Kepler-11 e & 0.0364 & 39.48 & 88.89 & 31.9996 & 6 & 40 \\
Kepler-11 g & 0.0290 & 94.40 & 89.87 & 118.3807 & 6 & 12 \\
Kepler-79 d & 0.0504 & 47.42 & 89.93 & 52.0902 & 6 & 25 \\
Kepler-89 d & 0.0680 & 23.80 & 89.97 & 22.3430 & 6 & 29 \\
Kepler-89 e & 0.0396 & 43.10 & 89.60 & 54.3203 & 4 & 10 \\
Kepler-90 g & 0.0615 & 127.30 & 89.89 & 210.6070 & 3 & 3 \\
Kepler-90 h & 0.0866 & 180.70 & 90.00 & 331.6006 & 1 & 1 \\
Kepler-103 c & 0.0335 & 98.00 & 89.70 & 179.6098 & 2 & 5 \\
Kepler-108 b & 0.0375 & 28.73 & 89.94 & 49.1839 & 3 & 8 \\
Kepler-108 c & 0.0330 & 70.81 & 89.40 & 190.3235 & 1 & 2 \\
Kepler-111 c & 0.0568 & 141.00 & 89.76 & 224.7783 & 2 & 4 \\
Kepler-199 b & 0.0295 & 37.08 & 89.97 & 23.6376 & 2 & 2 \\
Kepler-199 c & 0.0310 & 76.61 & 90.00 & 67.0934 & 1 & 1 \\
Kepler-209 c & 0.0371 & 21.43 & 87.50 & 41.7499 & 3 & 11 \\
Kepler-289 c & 0.1030 & 108.95 & 89.78 & 125.8723 & 5 & 5 \\
Kepler-396 b & 0.0302 & 72.95 & 90.00 & 42.9940 & 4 & 12 \\
Kepler-396 c & 0.0524 & 96.10 & 90.00 & 88.5050 & 4 & 5 \\
Kepler-411 b & 0.0266 & 9.86 & 87.40 & 3.0052 & 6 & 128 \\
Kepler-411 c & 0.0449 & 22.21 & 88.61 & 7.8344 & 5 & 49 \\
Kepler-411 d & 0.0372 & 73.40 & 89.43 & 58.0204 & 3 & 5 \\
Kepler-419 b & 0.0626 & 43.09 & 88.95 & 69.7546 & 4 & 11 \\
Kepler-450 b & 0.0346 & 26.38 & 88.82 & 28.4549 & 6 & 33 \\
Kepler-458 c & 0.0299 & 10.28 & 88.27 & 20.7403 & 2 & 16 \\
Kepler-468 b & 0.1400 & 66.25 & 90.00 & 38.4788 & 2 & 13 \\
Kepler-482 b & 0.0302 & 133.70 & 90.00 & 56.3542 & 4 & 13 \\
Kepler-511 b & 0.0484 & 132.39 & 89.81 & 296.6379 & 1 & 1 \\
Kepler-522 b & 0.0323 & 29.35 & 89.50 & 38.5842 & 6 & 25 \\
Kepler-530 b & 0.0312 & 53.61 & 89.97 & 39.3094 & 1 & 3 \\
Kepler-531 b & 0.0339 & 37.00 & 89.06 & 29.8849 & 2 & 4 \\
Kepler-533 b & 0.0432 & 40.00 & 88.68 & 28.5112 & 2 & 4 \\
Kepler-539 b & 0.0817 & 94.61 & 89.85 & 125.6324 & 4 & 4 \\
Kepler-803 b & 0.0310 & 61.77 & 89.98 & 50.2864 & 1 & 1 \\
Kepler-820 b & 0.0410 & 43.80 & 88.90 & 127.8338 & 1 & 1 \\
Kepler-858 b & 0.0493 & 76.79 & 89.97 & 76.1360 & 2 & 4 \\
Kepler-1311 d & 0.0535 & 90.20 & 89.49 & 197.5110 & 1 & 1 \\
Kepler-1656 b & 0.0419 & 38.70 & 89.31 & 31.5620 & 4 & 12 \\
Kepler-1662 b & 0.0710 & 94.80 & 89.44 & 134.4785 & 1 & 1 \\
Kepler-1731 b & 0.0322 & 137.07 & 89.95 & 21.7613 & 1 & 2 \\
Kepler-1746 b & 0.0514 & 377.00 & 89.87 & 321.2120 & 1 & 1 \\
Kepler-1868 b & 0.0512 & 104.70 & 89.50 & 211.0340 & 1 & 1 \\
KOI-00075.01 & 0.0381 & 39.70 & 89.22 & 105.8813 & 7 & 14 \\
KOI-00130.01 & 0.1126 & 44.43 & 89.19 & 34.1936 & 1 & 11 \\
KOI-00319.01 & 0.0467 & 29.37 & 88.24 & 46.1511 & 4 & 8 \\
KOI-00353.01 & 0.0640 & 93.78 & 89.46 & 152.1060 & 1 & 1 \\
KOI-00366.01 & 0.0634 & 75.91 & 89.37 & 75.1122 & 1 & 7 \\
KOI-00683.01 & 0.0517 & 314.00 & 89.84 & 278.1244 & 1 & 1 \\
KOI-00686.01 & 0.1177 & 108.58 & 89.64 & 52.5135 & 1 & 1 \\
KOI-00716.01 & 0.0683 & 37.60 & 88.51 & 26.8931 & 1 & 1 \\
KOI-01242.01 & 0.0720 & 69.73 & 89.21 & 99.6443 & 1 & 1 \\
KOI-01271.01 & 0.0693 & 104.64 & 89.64 & 162.0516 & 1 & 1 \\
KOI-02758.01 & 0.0390 & 34.89 & 89.98 & 253.3550 & 1 & 1 \\
\bottomrule
\end{tabular}
\\[1ex] 
\begin{minipage}{\columnwidth}
\small
Columns \#~Seg. and \#~Tr. indicate the number of segments and transits inspected for each planet.
\end{minipage}
\end{table}

\subsection{Simulated Sample}

\subsubsection{Simulation Code}

To evaluate the detection efficiency of the collaborators, we developed a Python simulation code capable of modeling complex configurations beyond the standard transit model. The software computes the stellar flux during planetary transits, incorporating stellar surface spots, exomoons, and TTVs to generate realistic synthetic light curves.

The routine generates a two-dimensional intensity map of the stellar disk with a configurable resolution. A quadratic limb darkening model is applied on this plane, which determines the intensity of each stellar region based on its distance to the center of the disk. This intensity map is the basis for calculating the flux occulted by any transiting body or structure over the stellar disk.

The code simulates stellar spots characterized by a parameterized number, size, and contrast. While the intrinsic properties of the spots remain static over time, they are carried by the stellar rotation, which alters their longitudinal position and modulates their visibility across the stellar hemisphere. To maximize the probability of SSCEs, we restricted the spots to equatorial latitudes. During a transit, the software evaluates the overlap between the planetary disk and the spotted regions. If a crossing occurs, we compute the resulting flux increment, caused by the lower intensity of the spot, by directly adding the flux of the occulted pixels on the stellar map.

Additionally, the simulation models the dynamics of planet-satellite systems. For simplicity, we assume circular orbits for both the planet and its satellite. The algorithm computes the relative positions of both bodies as a function of time, integrating the total occulted flux whenever they transit the stellar disk. To guarantee the occurrence of primary transits, satellite transits, and mutual planet-satellite eclipses, we fixed the orbital inclinations of both bodies to $i=90^\circ$.

Furthermore, the simulation incorporates TTVs by modulating the transit epochs with a sinusoidal function. Although this simplified approach does not alter the individual transit morphology (e.g., associated TDVs are not modeled), we implemented it to evaluate the robustness of our fitting algorithm. This ensures that the subsequent modeling procedure accurately recovers transit mid-times even when they deviate significantly from a linear ephemeris.

\begin{figure}[ht]
\centering
\includegraphics[width=0.75\columnwidth]{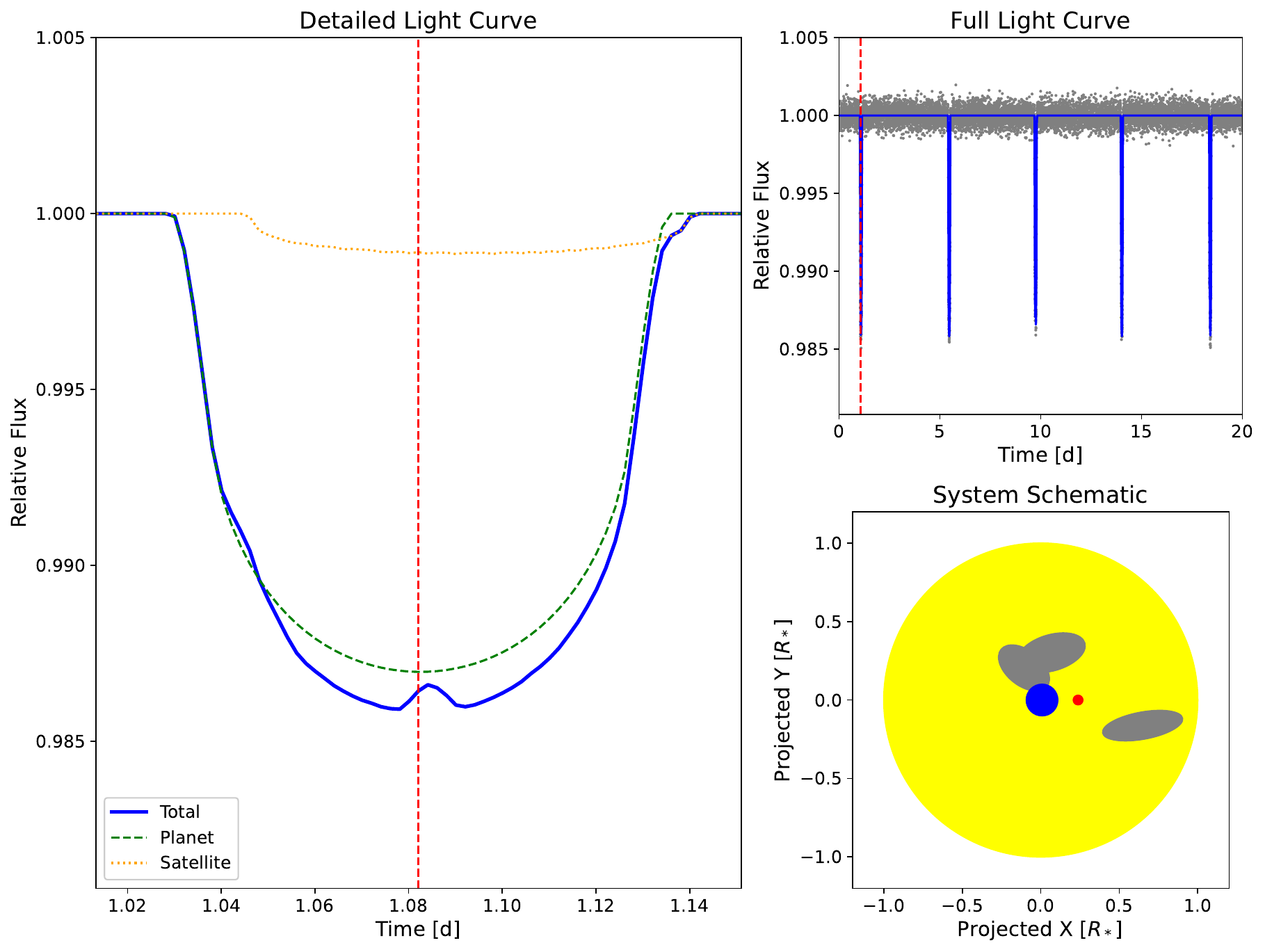}
\caption{Simulated light curve and system geometry. Example of a 20 d synthetic light curve of a solar-type star
($R_{\star} = 1 R_{\odot}$) featuring a transiting exoplanet ($R_{\rm p}/R_{\star} = 0.1$), an
exomoon ($R_{\rm s}/R_{\rm p} = 0.3$), and three stellar spots. Left panel: Detailed
view showing the individual contributions from the planet
(green dashed line) and the satellite (orange dotted line) to the
total flux (blue line). The bump in the total flux observed at the
center of the transit corresponds to an SSCE event. Upper right
panel: Full 20-day light curve with injected noise. Lower right
panel: 2D projection of the system at t = 1.08 d.}
\label{Fig:simulation}
\end{figure}

Finally, we injected Gaussian white noise ranging from 50 to 800 ppm into the resulting theoretical flux to reproduce the typical photometric precision of the selected \textit{Kepler} light curves. Additionally, we applied a temporal offset to align the simulated transit epochs with the time format of the archival \textit{Kepler} data. In Fig.~\ref{Fig:simulation} we show an example of a synthetic light curve generated with the code.

\subsubsection{Simulated light curves}

Using the simulation framework described above, we generated 100 synthetic light curves containing planetary transits, strictly reproducing the observational characteristics of the original \textit{Kepler} mission. Specifically, we adopted a sampling cadence of 1~min to emulate the \textit{Kepler} short-cadence (SC) observation mode. To account for the finite integration time of \textit{Kepler}, we computed the model at a higher temporal resolution using a supersampling factor of 10, subsequently averaging these points to match the 1~min sampling cadence. For each light curve, spanning a baseline of 100~d, we drew the stellar mass from a uniform distribution between 0.2 and 1.5~$M_{\odot}$, and estimated the stellar radius using the mass-radius relation $R_{\star} = (M_{\star}/M_{\odot})^{0.8} R_{\odot}$. We assumed constant quadratic limb-darkening coefficients of $u_1=0.6$ and $u_2=0.2$. Next, we sampled the planetary orbital period, $P_{\rm orb}$, between 20 and 50~d. We adopted the lower limit of 20~d to allow for the dynamical stability of potential exomoons, and imposed the upper limit of 50~d to guarantee the occurrence of at least two transits within the observational window. After drawing the planetary radius between 1 and 10~$R_{\oplus}$, we derived the orbital semi-major axis from $P_{\rm orb}$ and $M_{\star}$ using Kepler's third law.

We generated 50 synthetic light curves corresponding to the null case (i.e., containing only standard planetary transits). For these simulations, we randomly sampled the stellar and planetary parameters and injected photometric noise ranging from 50 to 800~ppm, which is consistent with \textit{Kepler} targets brighter than $K_{\rm p} = 14$. Furthermore, to verify the capability of our pipeline to accurately recover signals in the presence of timing deviations, we injected TTVs into 10 of these light curves, applying amplitudes of 15~min (five curves) and 5~min (five curves).

Additionally, we generated 40 synthetic light curves incorporating stellar spots to model SSCEs. To evaluate the visual detection threshold of the collaborators, we classified the configurations into three detectability tiers---easy, medium, and difficult---driven by the fractional area of the spots and the injected photometric noise. Specifically, we simulated 15 curves for the easy tier, featuring large spots ($0.05 \le R_{\rm spot}/R_{\star} \le 0.10$) in low-noise conditions (50--100~ppm). Conversely, the difficult tier comprised 10 curves with smaller spots ($0.02 \le R_{\rm spot}/R_{\star} \le 0.05$) and high noise (400--800~ppm), while the remaining 15 curves represented an intermediate medium tier.

Finally, we generated 10 synthetic light curves modeling transits of planet-satellite systems. To ensure the exomoons were visually detectable by the collaborators, we fixed the planetary radius to $R_{\rm p} = 1~R_{\rm Jup}$ and the satellite radius to $R_{\rm s} = 0.25~R_{\rm Jup}$. To guarantee dynamical stability, we restricted the satellite's orbital semi-major axis to values between 0.1 and 0.4 times the planetary Hill radius. Additionally, we injected photometric noise ranging from 50 to 200~ppm into these simulations.

\subsection{Transit analysis}

The methodology presented in this work is based on a systematic comparison between individual transit events and a global template. For each real and simulated light curve, we construct a theoretical model representing the average transit profile, derived from the known stellar and planetary parameters of the system. This global model effectively marginalizes over the stochastic variations of individual events. By superimposing this average profile onto each individual transit, we can identify significant residuals that may indicate the presence of transient phenomena or additional signals.

\subsubsection{Global transit modeling and optimization}

By using the \texttt{batman} package \citep{2015ascl.soft10002K}, our analysis pipeline generates theoretical transit models based on the formalism of \citet{2002ApJ...580L.171M}. For each light curve, the code explores a grid of planetary radii and semi-major axes within a $\pm15\%$ range of the values provided in the metadata of the light curves. To minimize parameter degeneracy, particularly between the orbital inclination and the semi-major axis, and to optimize computational performance across the dataset, we opted to fix the inclination to its nominal value during the fitting process.

\begin{figure}[t]
\centering
\includegraphics[width=0.82\columnwidth]{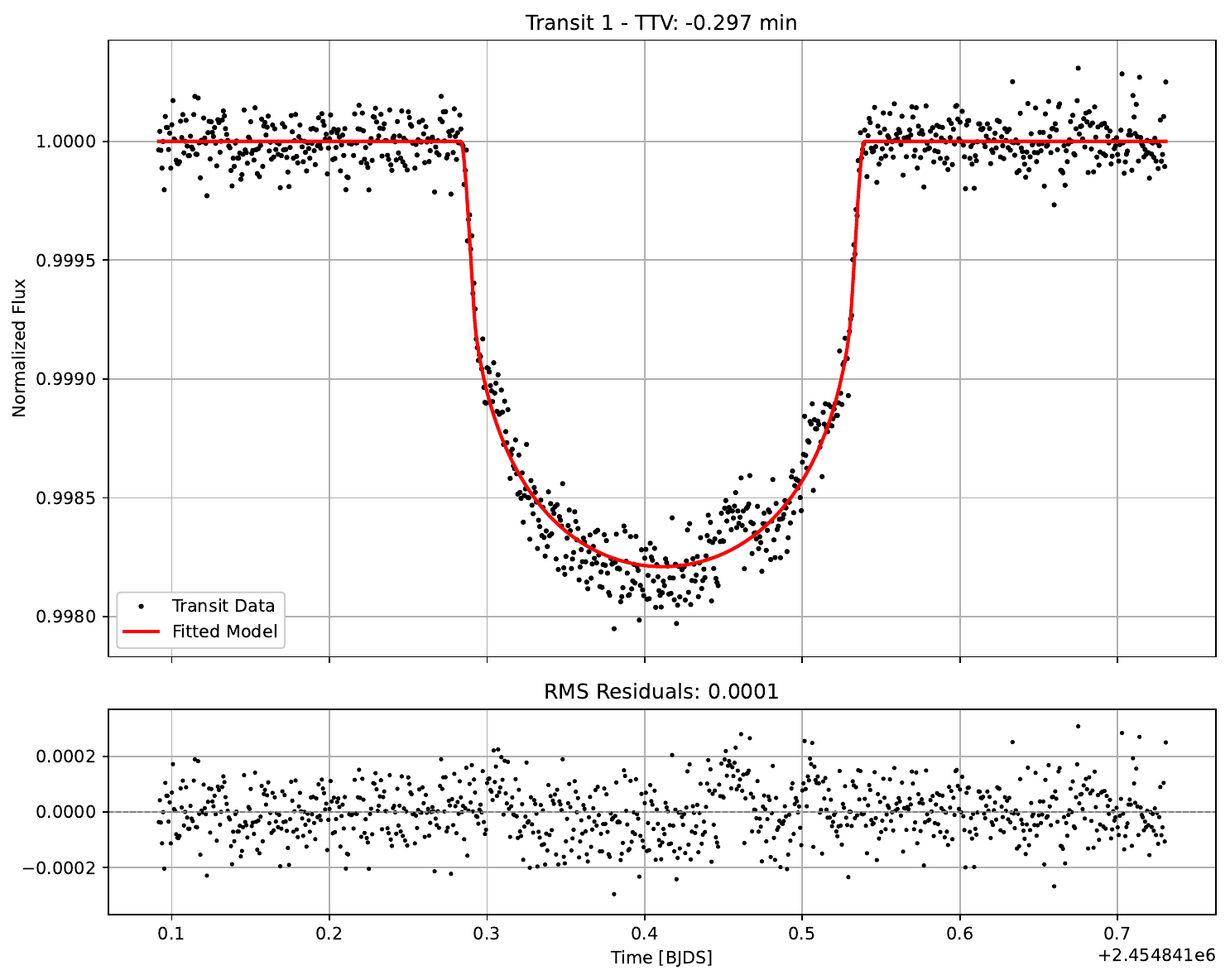}

\vspace{0.5cm}

\includegraphics[width=0.82\columnwidth]{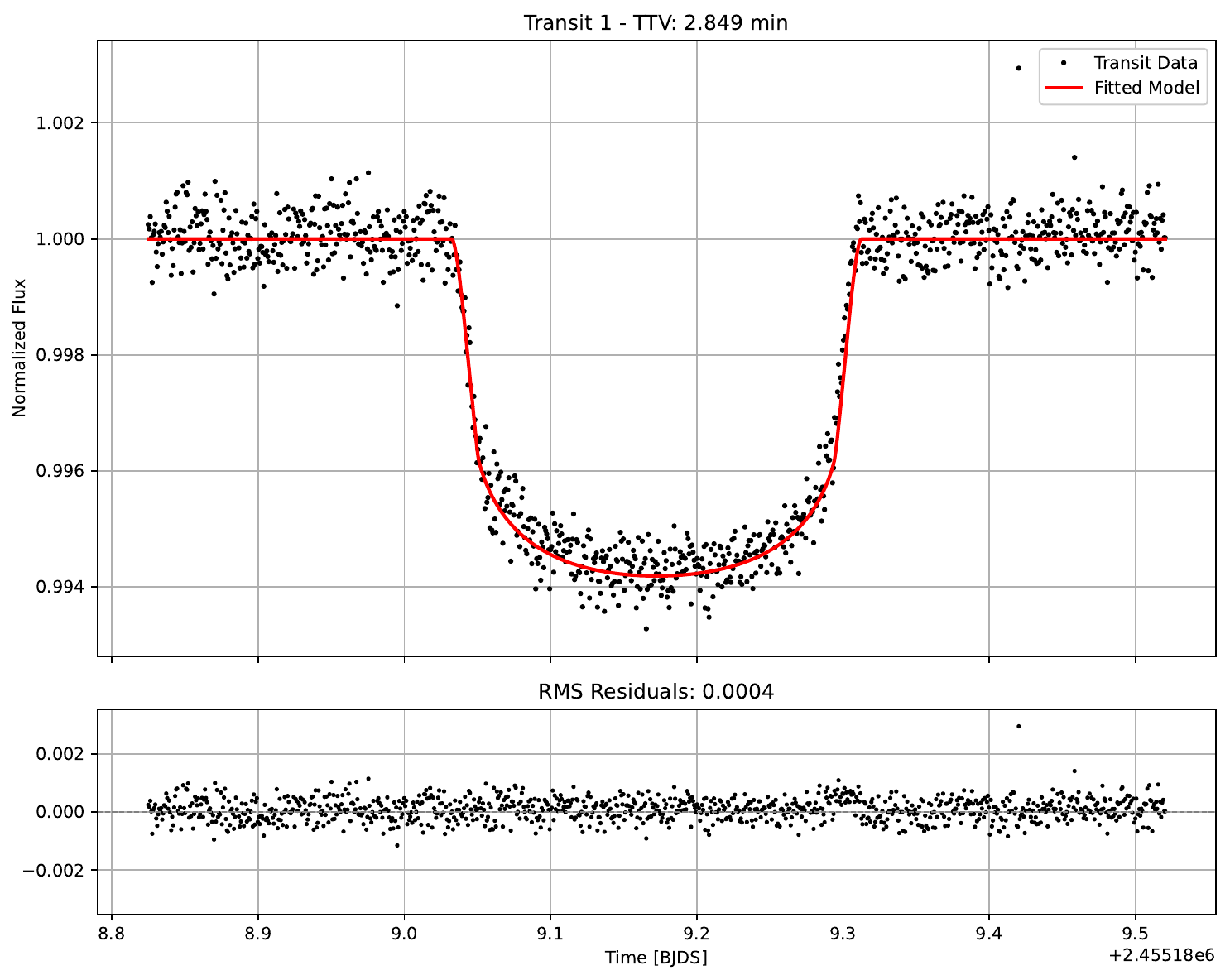}

\caption{Examples of the model fit. Simulated (top panel) and real (bottom panel, Kepler-89 d).}
\label{Fig:Example_transits}
\end{figure}

For each $(R_{\rm p}, a)$ pair, the algorithm iterates through all individual transits to determine their precise temporal midpoints. To account for potential TTVs, the code employs a time-adaptive search window initiated from five temporal seeds. These seeds are distributed across a range proportional to the transit duration and centered on the linear ephemeris of each event. The algorithm computes the $\chi^{2}$ statistic for each transit, yielding a total $\chi^{2}$ value for the entire light curve associated with that specific $(R_{\rm p}, a)$ parameter set. We then find the optimal global fit parameters by minimizing the total $\chi^{2}$ using the L-BFGS-B algorithm \citep{1995SJSC...16.1190B}. The optimizer was executed with its default numerical configuration; its conservative step size, in conjunction with the strict linear ephemerides, naturally constrained large fluctuations in the scaled semi-major axis ($a/R_{\star}$) during this initial screening phase. This resulting global model serves as the template to be superimposed onto each transit during the subsequent visual inspection phase. In Fig.~\ref{Fig:Example_transits} we show examples of the model fit of our code to both, real and simulated data.

\subsubsection{Classification platform and collaborator training}

The analysis code is accessed via a dedicated web interface, allowing users to inspect all available light curves and efficiently navigate through every transit event. Each transit is visualized with the previously described global theoretical model superimposed, alongside the corresponding fit residuals. The tool also tracks the user's progress, displaying the number of light curves and individual transits analyzed out of the total dataset.

For each transit, users utilize a set of checkboxes to flag specific morphological features indicative of the phenomena under investigation. The available classification categories are: normal morphology, anomalous morphology, left asymmetry, right asymmetry, interior flux increase, interior flux decrease, transit duration variation, poor model fit, and a free-text field for additional notes. All classifications are recorded in a centralized database for subsequent analysis.

Given the manageable size of the dataset (240 light curves comprising 917 transits), we opted for a comprehensive approach where every collaborator independently reviewed all available transits. The inspection was conducted blindly; users were unaware of whether a given light curve was real or simulated. The collaborators are experienced amateur astronomers, many of whom previously participated in the K2-OjOS project inspecting light curves for exoplanetary signals. Several members also possess an extensive background in astronomical observation (particularly photometry and astrometry) and prior experience in pro-am collaborations. To ensure classification consistency during this pilot phase, we provided detailed tutorials illustrating reference examples of the various expected anomalies, supplemented by periodic group meetings throughout the inspection period. The web-based classification tool, comprehensive project documentation, and the complete dataset of light curves (both real and synthetic) are publicly available at \url{https://dipsojos.uniovi.es}.

\subsubsection{Validation framework and collaborator weighting}

To evaluate the consistency and reliability of the collaborator's classifications, we implemented a validation protocol using the simulated light curves as a control sample. We defined a True Positive (TP) as any instance where a user marked anything other than ``normal morphology'' in a transit containing an injected signal, and a False Negative (FN) if the user marked ``normal morphology''. Conversely, for the null-case curves (containing a standard transit with noise), a True Negative (TN) was recorded if ``normal morphology'' was selected, and a False Positive (FP) if any other category was marked.

\begin{figure}[ht]
\centering
\includegraphics[width=0.8\columnwidth]{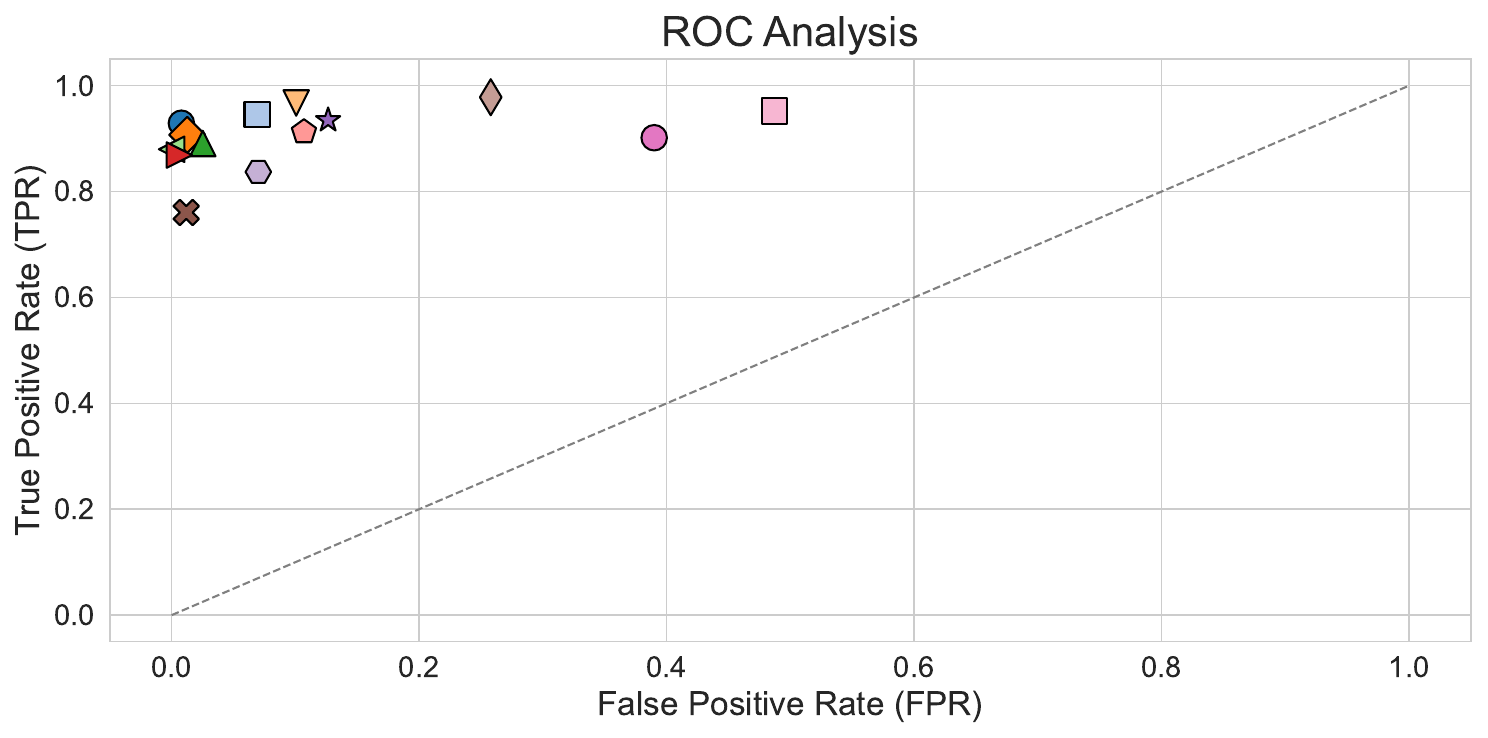}
\caption{Distribution of collaborators in the ROC space. Different symbols and colors identify individual observers.}
\label{GraficaROC}
\end{figure}

Using these definitions, we constructed a Receiver Operating Characteristic (ROC) curve for the collaborators (see Fig.~\ref{GraficaROC}). The observer is positioned on the Y-axis according to their True Positive Rate (TPR), defined as $TPR = \frac{TP}{TP+FN}$, which indicates their efficiency in recovering signals, so that collaborators located closer to $Y=1$ are more efficient at recovering injected signals. The $X$-axis represents the False Positive Rate (FPR), defined as $FPR = \frac{FP}{FP+TN}$, and reflecting their tendency to generate false alarms, so observers located closer to $X=0$ generate fewer false positives

The Area Under the Curve (AUC) in the ROC space serves as a metric of individual efficiency. Based on their AUC, collaborators were classified into three distinct categories: A ($\mathrm{AUC} > 0.8$), B ($0.7 < \mathrm{AUC} \leq 0.8$), and C ($\mathrm{AUC} \leq 0.7$). This categorization was subsequently used to weight the classifications performed on real transits, thereby reducing the total number of events requiring subsequent expert review from ICTEA researchers.

We established a filtering criterion requiring that a real transit must be flagged as anomalous by at least one type A classifier, two type B classifiers, or a combination of one type B and two type C classifiers. For the transits passing this filter, a final ranking score was computed by awarding 10 points for every type A classifier that marked an anomaly, 5 points for every type B, and 1 point for every type C.

\section{Results and discussion}

\subsection{Simulated control sample}

The analysis of the simulated control dataset provided a robust baseline for the efficiency of the visual inspection. For the null-case scenario, which comprised 50 light curves and a total of 159 individual transits, the collaborators yielded an average FPR of 15.3\%.

Regarding the recovery of injected signals, the participants demonstrated high sensitivity. In the subset of light curves injected with starspots (40 curves comprising 151 transits), the collaborators flagged the event as having a non-normal morphology in an average of 87.6\% of the cases. Furthermore, the ``interior flux increase'' checkbox, which constitutes the most prominent visual signature of a SSCE, was specifically selected in 64.9\% of these transits. Similarly, for the light curves injected with satellite signals (10 curves comprising 52 transits), the average recovery rate was 93.1\%.

Considering the entire sample of 203 injected signals (encompassing both starspots and satellites), the collective performance of the participants yielded a mean AUC of 0.876. This value indicates a high diagnostic accuracy, significantly exceeding the random classification threshold ($\mathrm{AUC} = 0.5$). Such a high degree of discriminative power confirms that the morphological criteria employed by the collaborators are robust, making their weighted consensus a highly reliable tool for prioritizing exoplanetary anomalies in archival data.

\subsection{Real sample}

Following the established methodology, after inspecting a sample of 555 real \textit{Kepler} transits we ended with 363 transits ($65\%$) being considered by the collaborators as ``suspicious''. This final subset, representing a manageable volume of data, was subsequently subjected to an expert visual inspection. The expert review reduced the sample to 88 viable candidates, categorized as follows: seven candidates for stellar flares, 41 for SSCEs, 38 with potential TDVs, and two exhibiting overlapping transits.

The transition from simulated control data to real observations highlights the emergence of FP and FN generators inherent to real data, which are absent in our simulated scenario, where FPs and FNs theoretically represent only human classification errors.
Based on the simulated null-case sample, the baseline FPR was established at 15.3\%. If real data were perfectly clean, applying this FPR and a TPR of $85\%$ to our 555 real transits, yielding 363 flagged candidates, would imply the presence of nearly 400 genuine astrophysical signals. However, the expert inspection retained only 88 viable candidates.

This discrepancy demonstrates that the effective FPR in real data is significantly higher than the 15.3\% reference measured in the simulations. We do not attribute this increase to a degradation in collaborator performance. Instead, it is due to the fact that real light curves contain artifacts, such as imperfect detrending in active stars, data gaps, and model fitting inaccuracies attributable to noise or incorrect theoretical parameters, that ultimately mimic the morphology of the astrophysical phenomena targeted in this study.

When collaborators encounter these artifacts, they correctly flag the transit as exhibiting an ``anomalous morphology''. Consequently, the 275 transits discarded during the expert review predominantly represent false positives inherent to the real data, rather than human errors.

We next present the signals recovered from the inspection of the real light curves. These candidates should be regarded as the first science-validation output of the DIPS-OjOS workflow rather than as an exhaustive catalog of transit anomalies. A comprehensive physical analysis of each individual event falls outside the scope of this pilot phase. Our primary objective is to validate the methodology, quantify the performance of the visual-inspection protocol, and establish the basis for future large-scale applications of the project.

\subsubsection{Starspot-crossing Events}
\label{sec:ssces}

The initial visual inspection by our collaborators followed by the expert revision yielded 41 signals classified as potential SSCEs. A subsequent analysis refined this sample, confirming 26 in-transit flux increases across the light curves of eight exoplanets as compatible with SSCEs. However, noise is present in the light curves, and detrending algorithms cannot always perfectly correct stellar variability. As a result, residual artifacts
that mimic the morphological signature of true SSCEs can emerge. To validate these visually identified signals, we implemented a protocol based on integrated area analysis. For each anomaly, we isolated its time window and fitted a univariate cubic spline (smoothing factor $s=10^{4}$). This regularization factor effectively filters high-frequency noise while preserving the SSCE structure, and provides an estimate of the event's central time.

To establish statistical confidence, we performed a local bootstrapping test with 10,000 iterations for each candidate. In each iteration, we fitted the same cubic spline to a control window of identical duration, randomly shifted across the available transit phase. By comparing the integrated area of the actual anomaly against the resulting empirical distribution of noise-generated areas, we derived a $p$-value that quantifies the probability of the signal being a purely stochastic artifact.

\begin{figure*}[ht]
\centering
\scriptsize

\begin{minipage}[t]{0.18\textwidth}
\centering
\includegraphics[width=\textwidth]{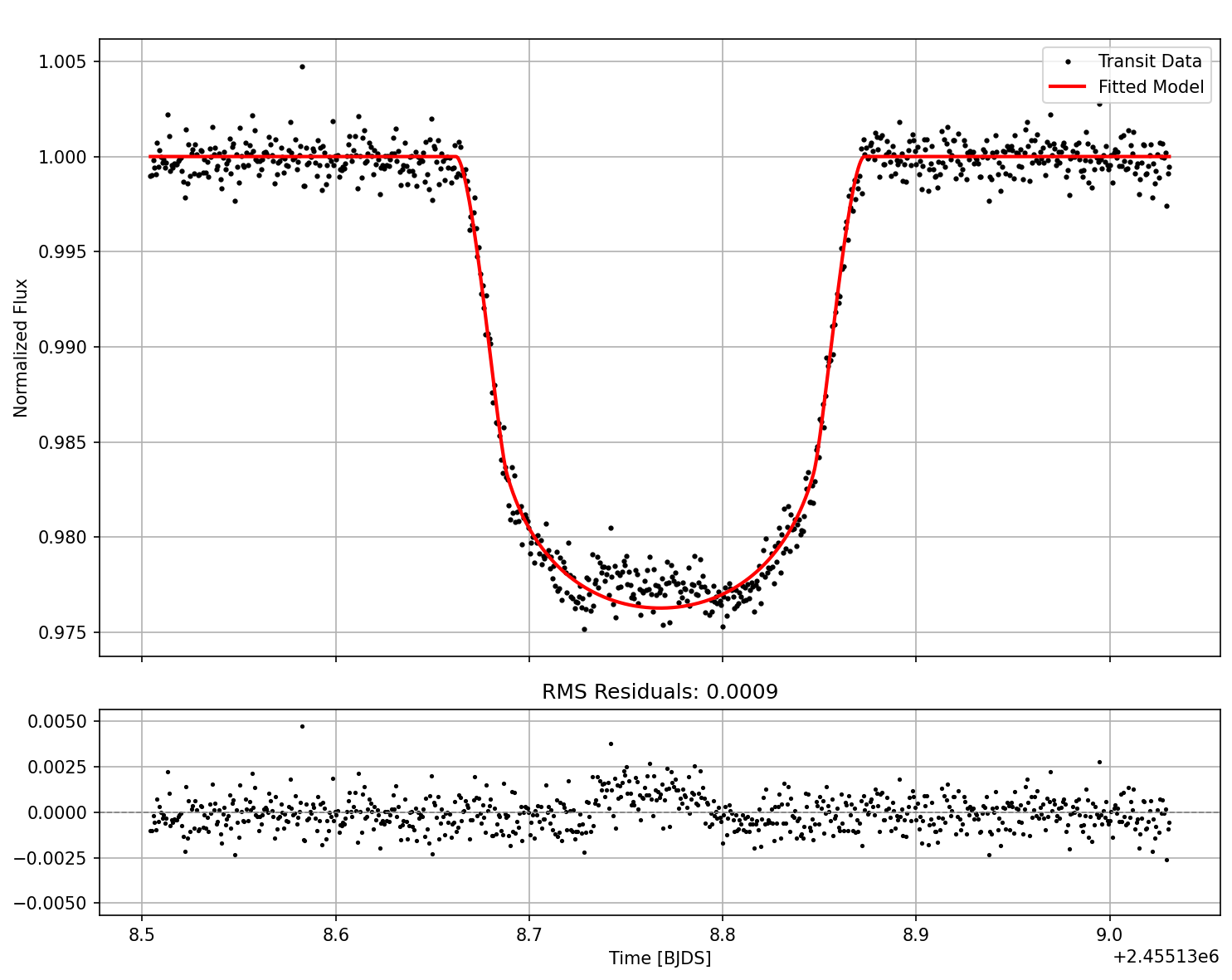}\\
Kepler-468 b (305.756)
\end{minipage}\hfill
\begin{minipage}[t]{0.18\textwidth}
\centering
\includegraphics[width=\textwidth]{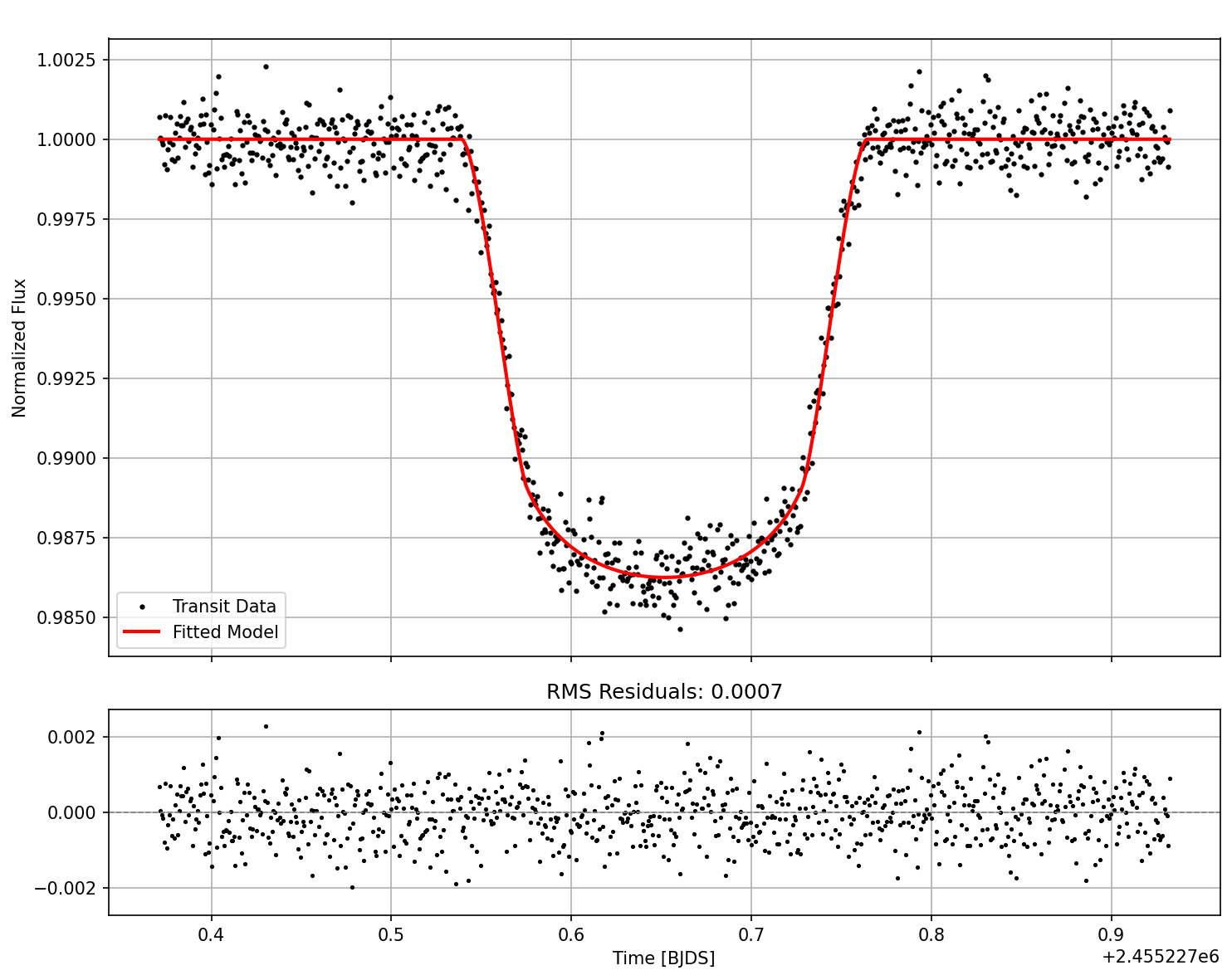}\\
KOI 00130.01 (394.671)
\end{minipage}\hfill
\begin{minipage}[t]{0.18\textwidth}
\centering
\includegraphics[width=\textwidth]{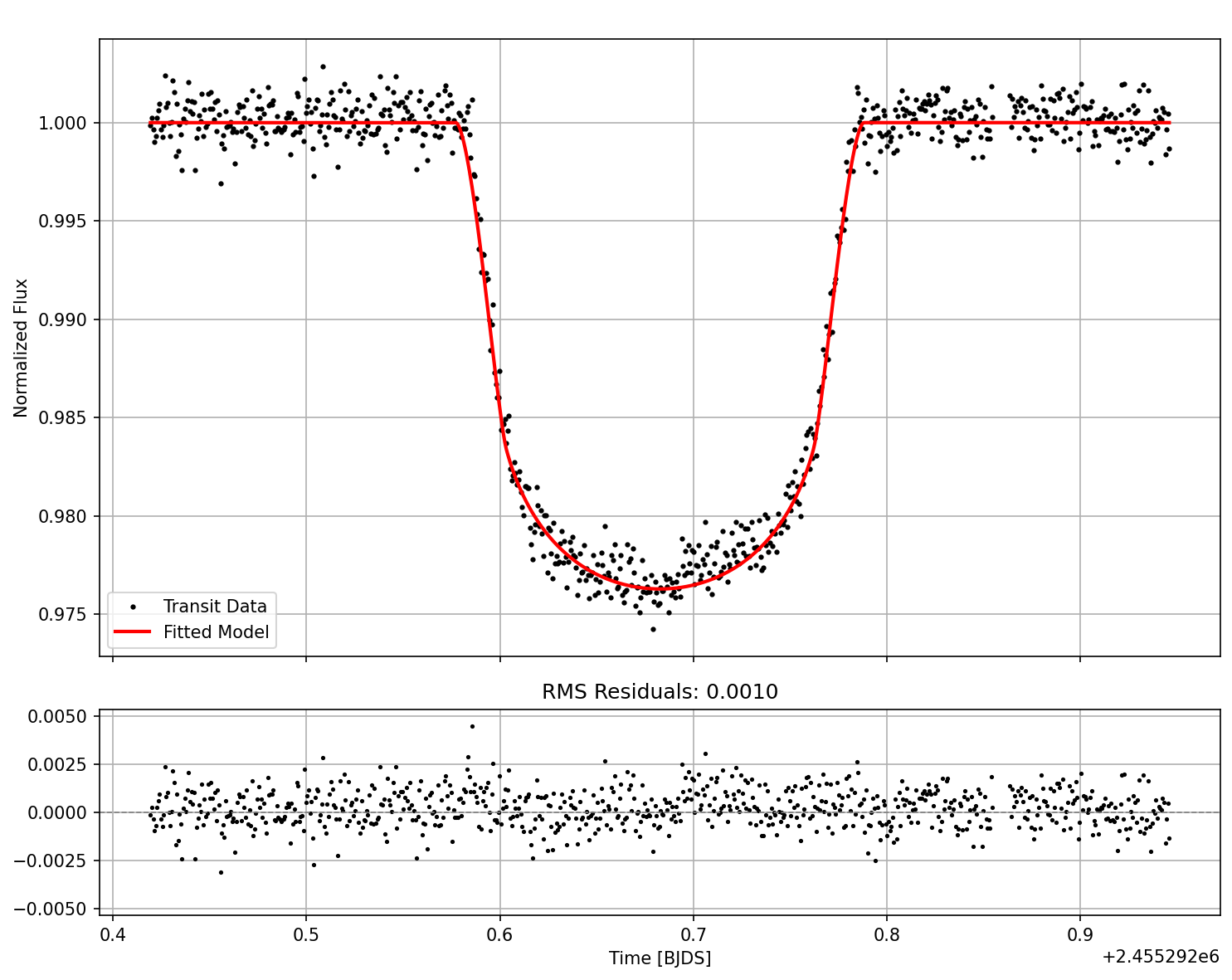}\\
Kepler-468 b (459.661 \& .706)
\end{minipage}\hfill
\begin{minipage}[t]{0.18\textwidth}
\centering
\includegraphics[width=\textwidth]{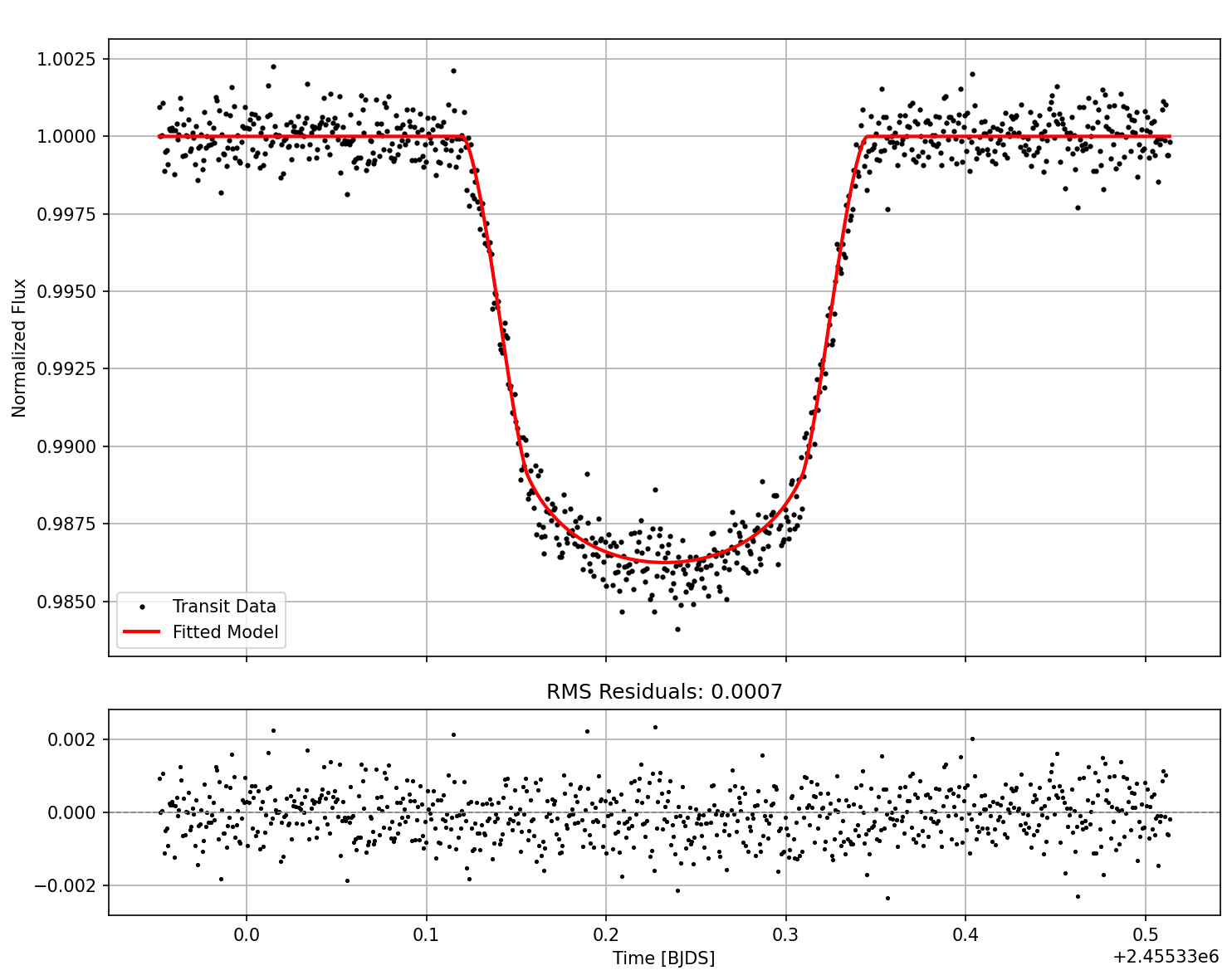}\\
KOI 00130.01 b (497.226)
\end{minipage}\hfill
\begin{minipage}[t]{0.18\textwidth}
\centering
\includegraphics[width=\textwidth]{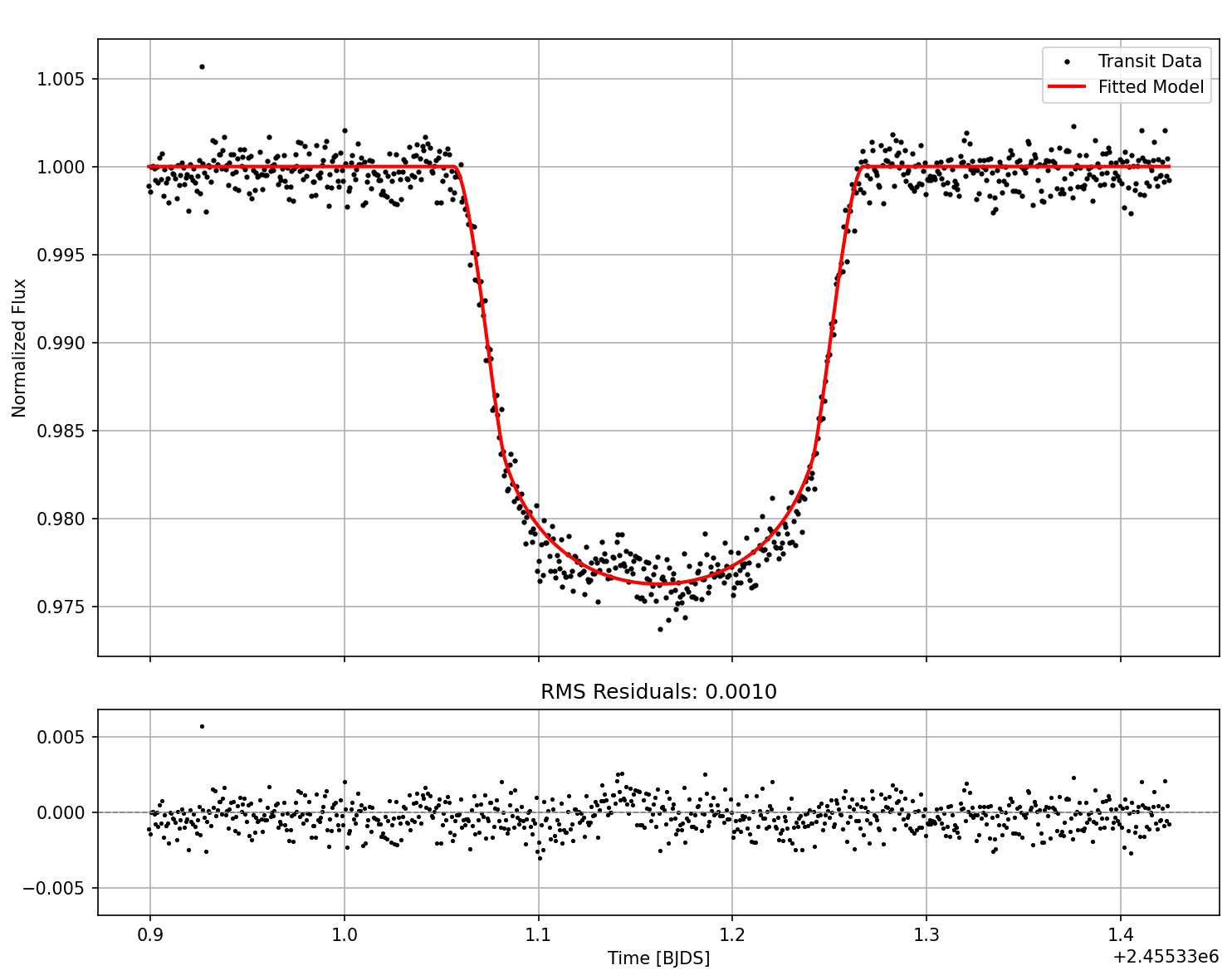}\\
Kepler-468 b (498.142)
\end{minipage}

\vspace{10pt}

\begin{minipage}[t]{0.18\textwidth}
\centering
\includegraphics[width=\textwidth]{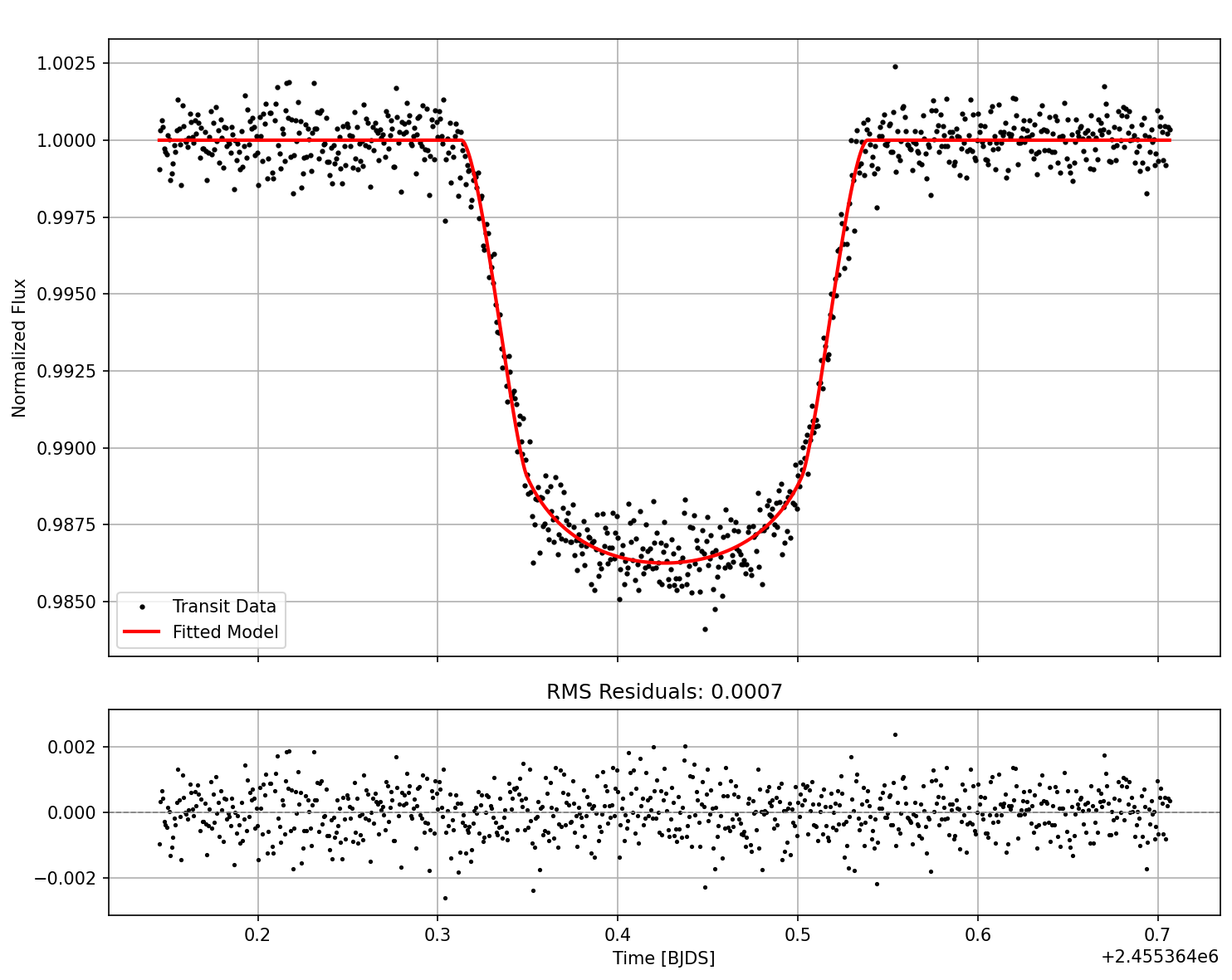}\\
KOI 00130.01 b (531.418)
\end{minipage}\hfill
\begin{minipage}[t]{0.18\textwidth}
\centering
\includegraphics[width=\textwidth]{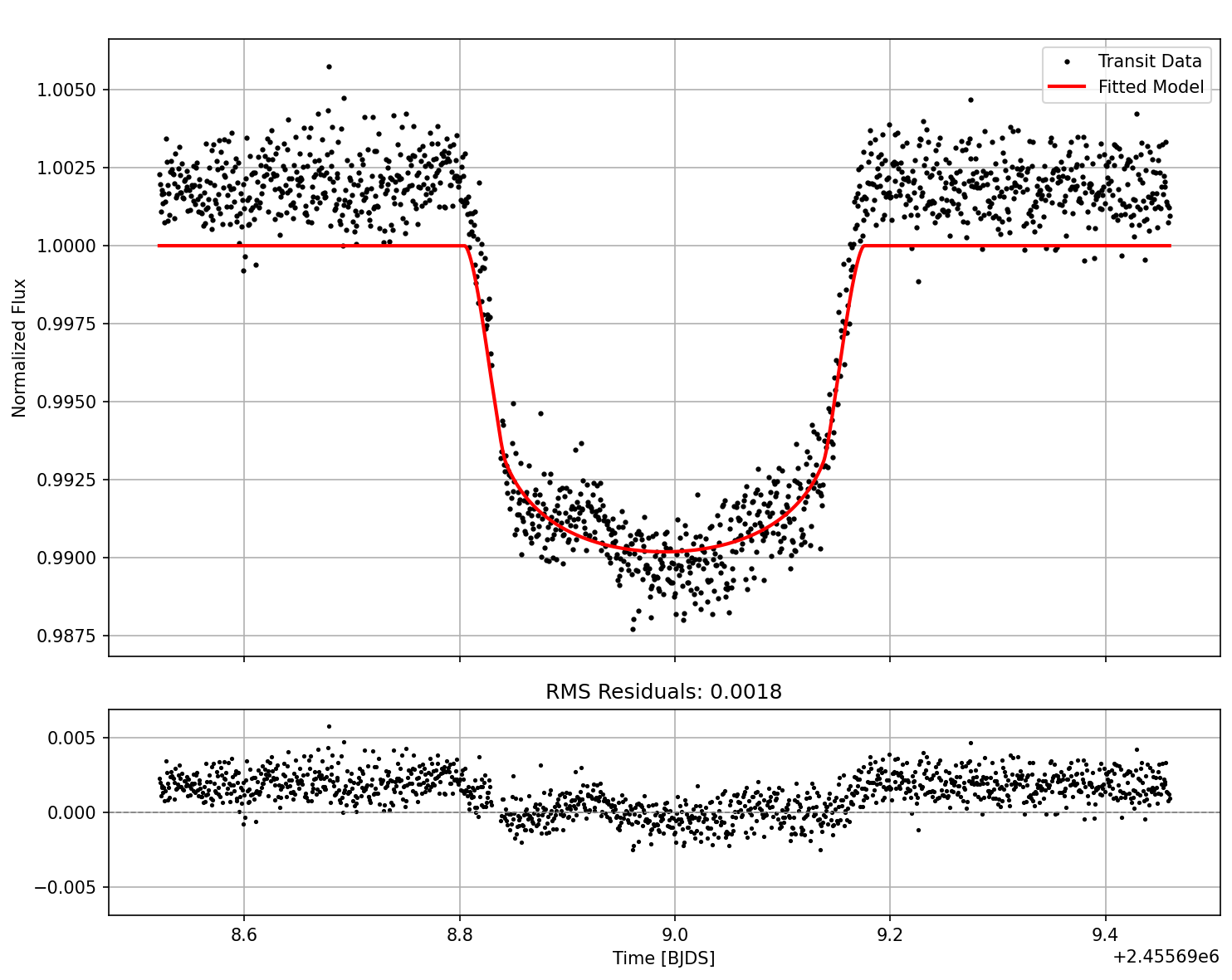}\\
Kepler-289 c (865.920)
\end{minipage}\hfill
\begin{minipage}[t]{0.18\textwidth}
\centering
\includegraphics[width=\textwidth]{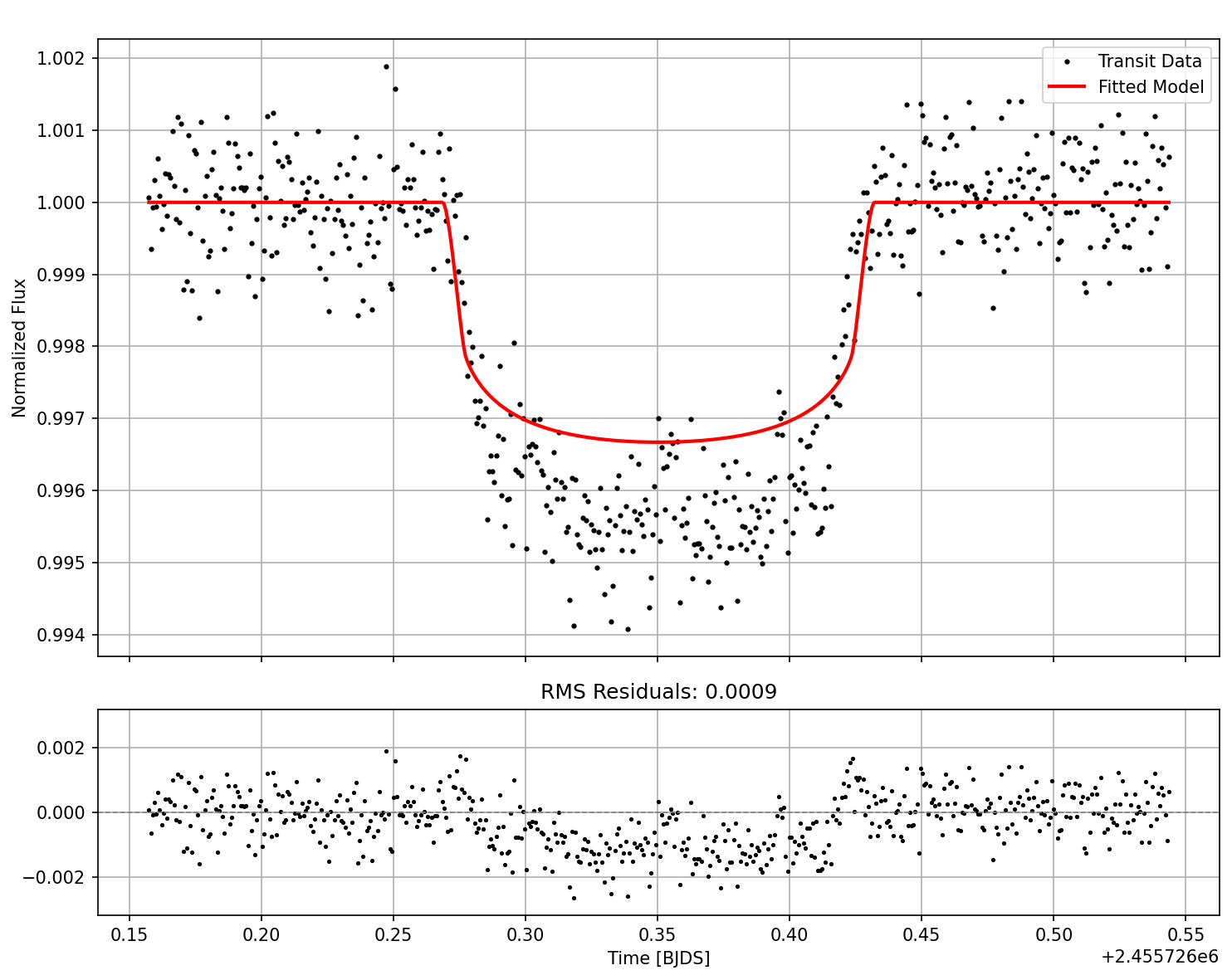}\\
Kepler-419b (893.353)
\end{minipage}\hfill
\begin{minipage}[t]{0.18\textwidth}
\centering
\includegraphics[width=\textwidth]{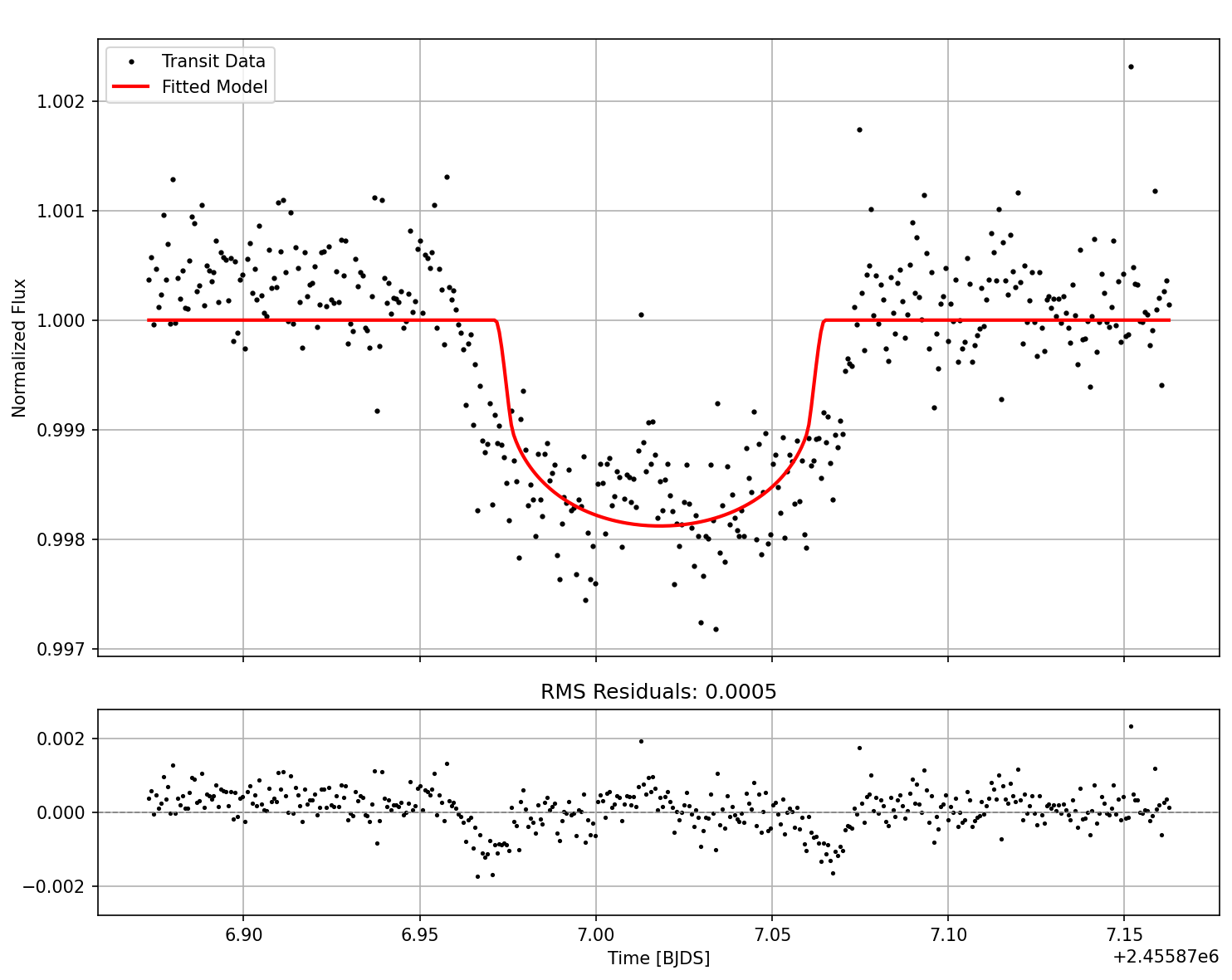}\\
Kepler-411 c (1044.013)
\end{minipage}\hfill
\begin{minipage}[t]{0.18\textwidth}
\centering
\includegraphics[width=\textwidth]{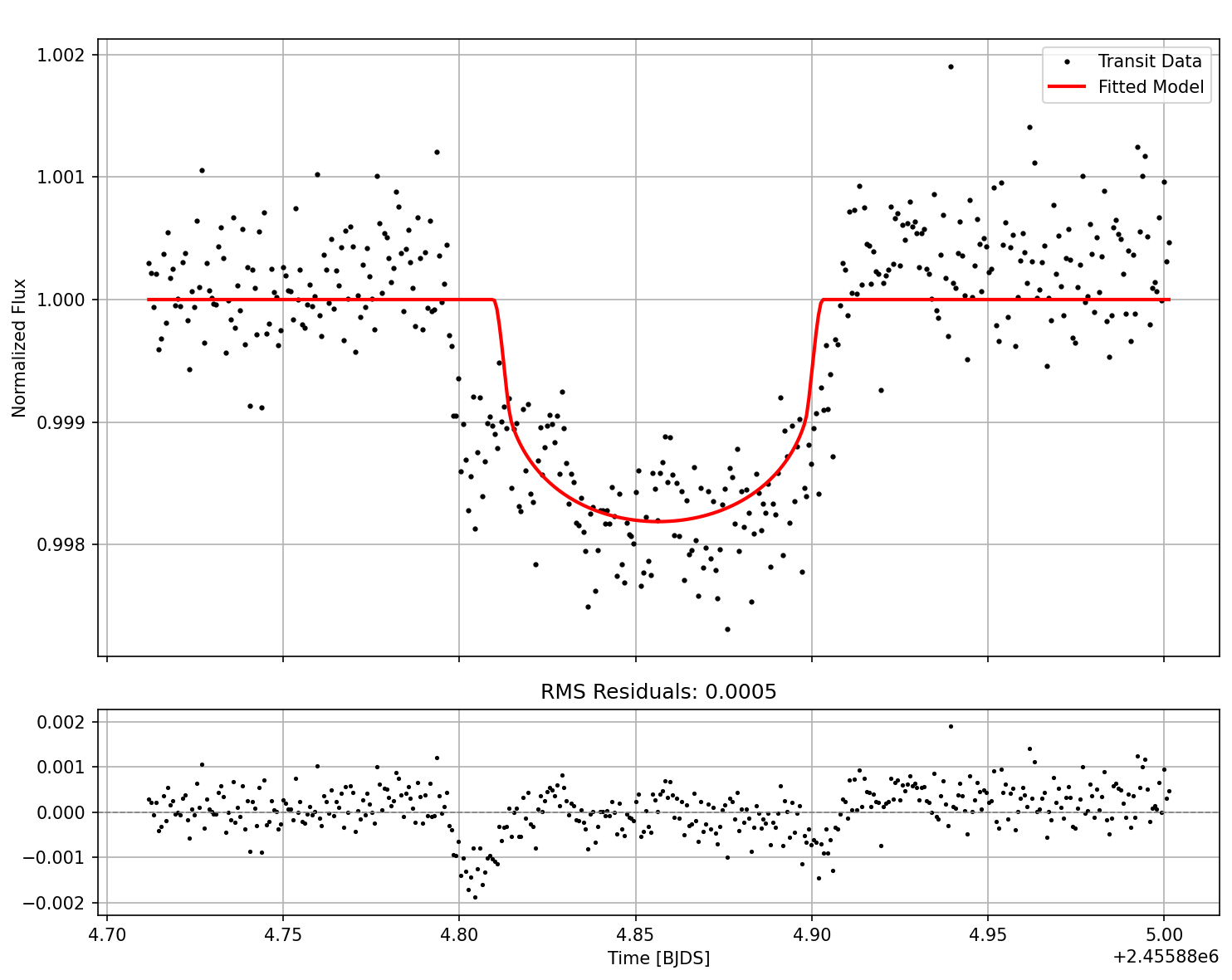}\\
Kepler-411 c (1051.825 \& .857)
\end{minipage}

\vspace{10pt}

\begin{minipage}[t]{0.18\textwidth}
\centering
\includegraphics[width=\textwidth]{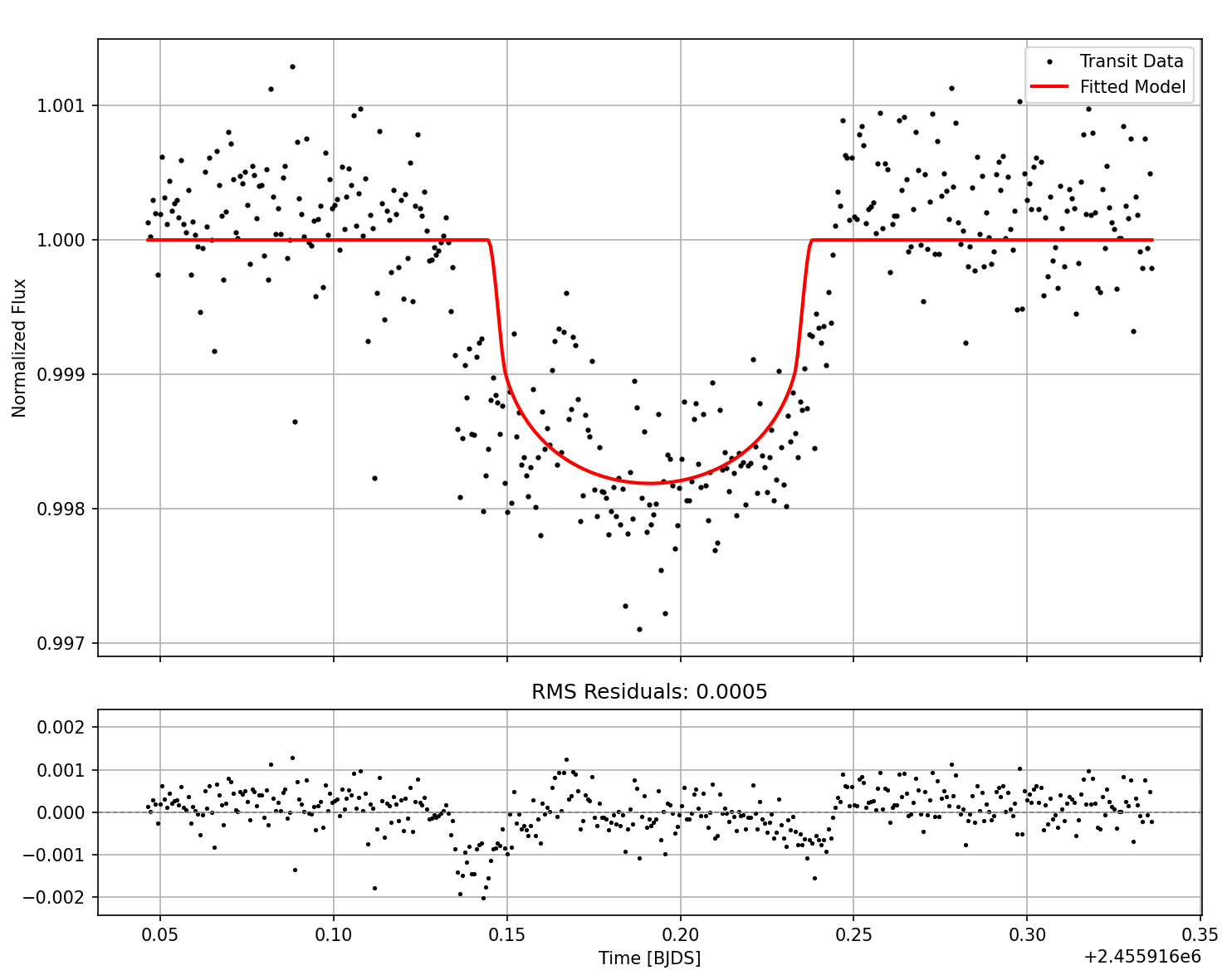}\\
Kepler-411 c (1083.166)
\end{minipage}\hfill
\begin{minipage}[t]{0.18\textwidth}
\centering
\includegraphics[width=\textwidth]{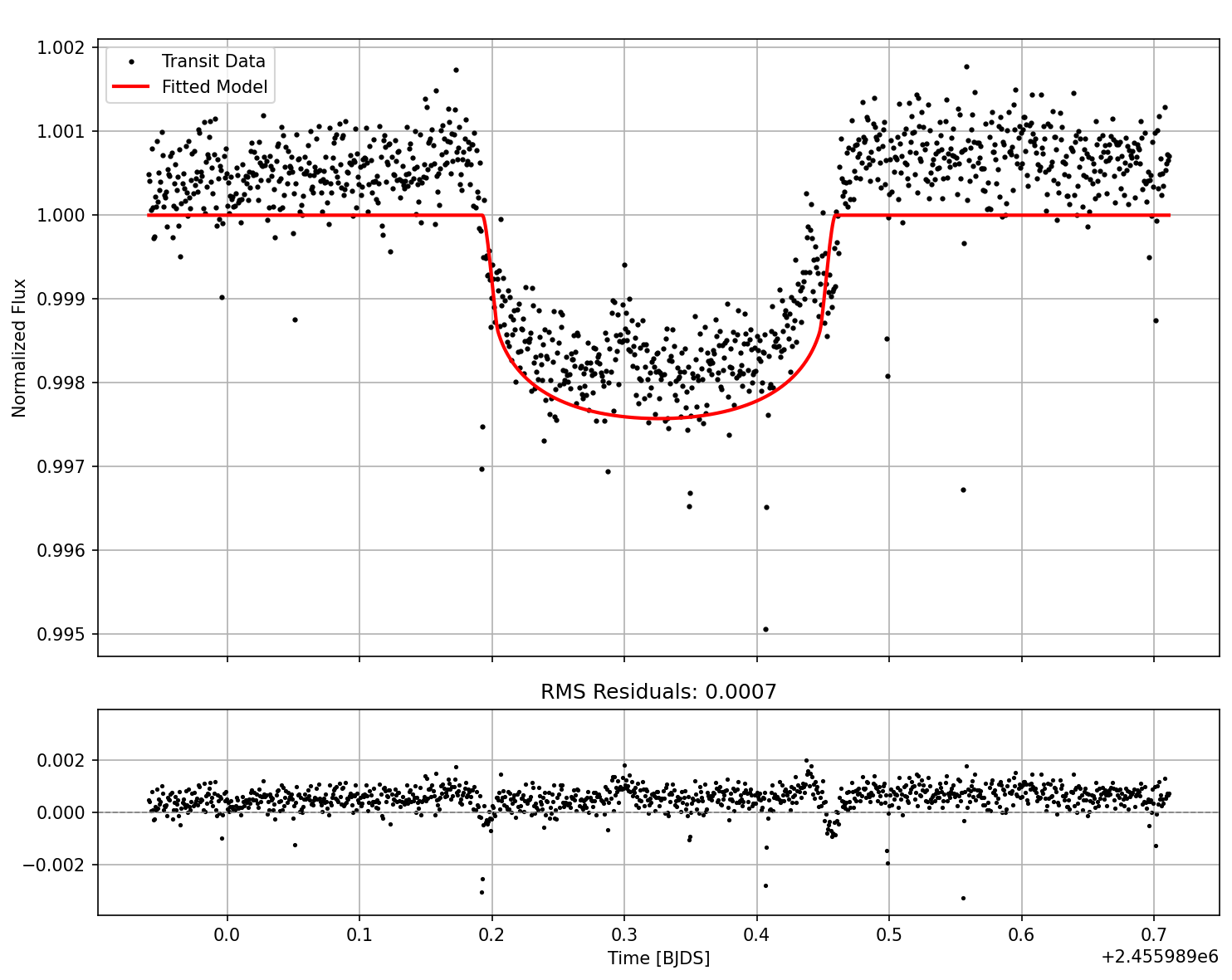}\\
Kepler-396 c (1156.297)
\end{minipage}\hfill
\begin{minipage}[t]{0.18\textwidth}
\centering
\includegraphics[width=\textwidth]{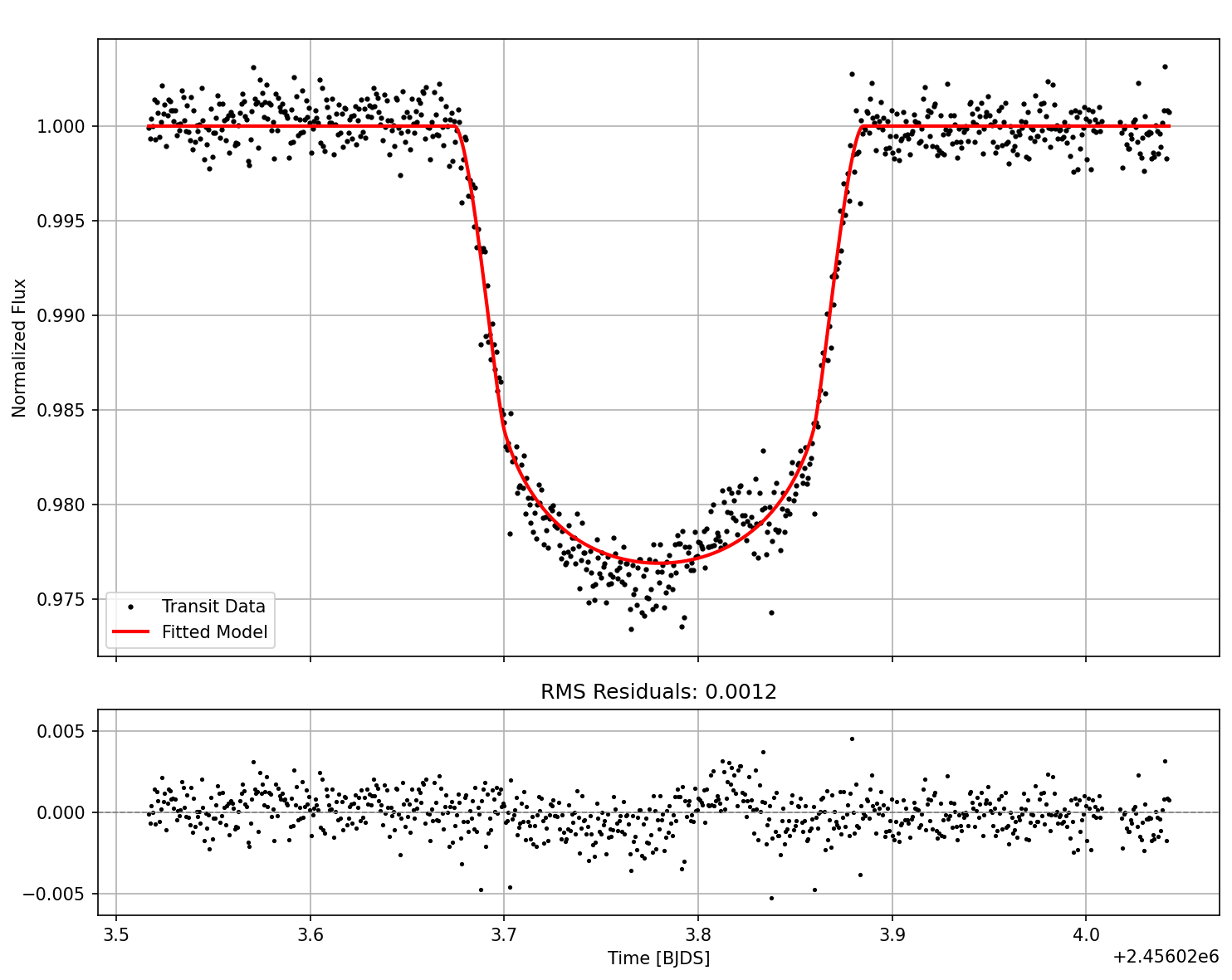}\\
Kepler-468 b (1190.816)
\end{minipage}\hfill
\begin{minipage}[t]{0.18\textwidth}
\centering
\includegraphics[width=\textwidth]{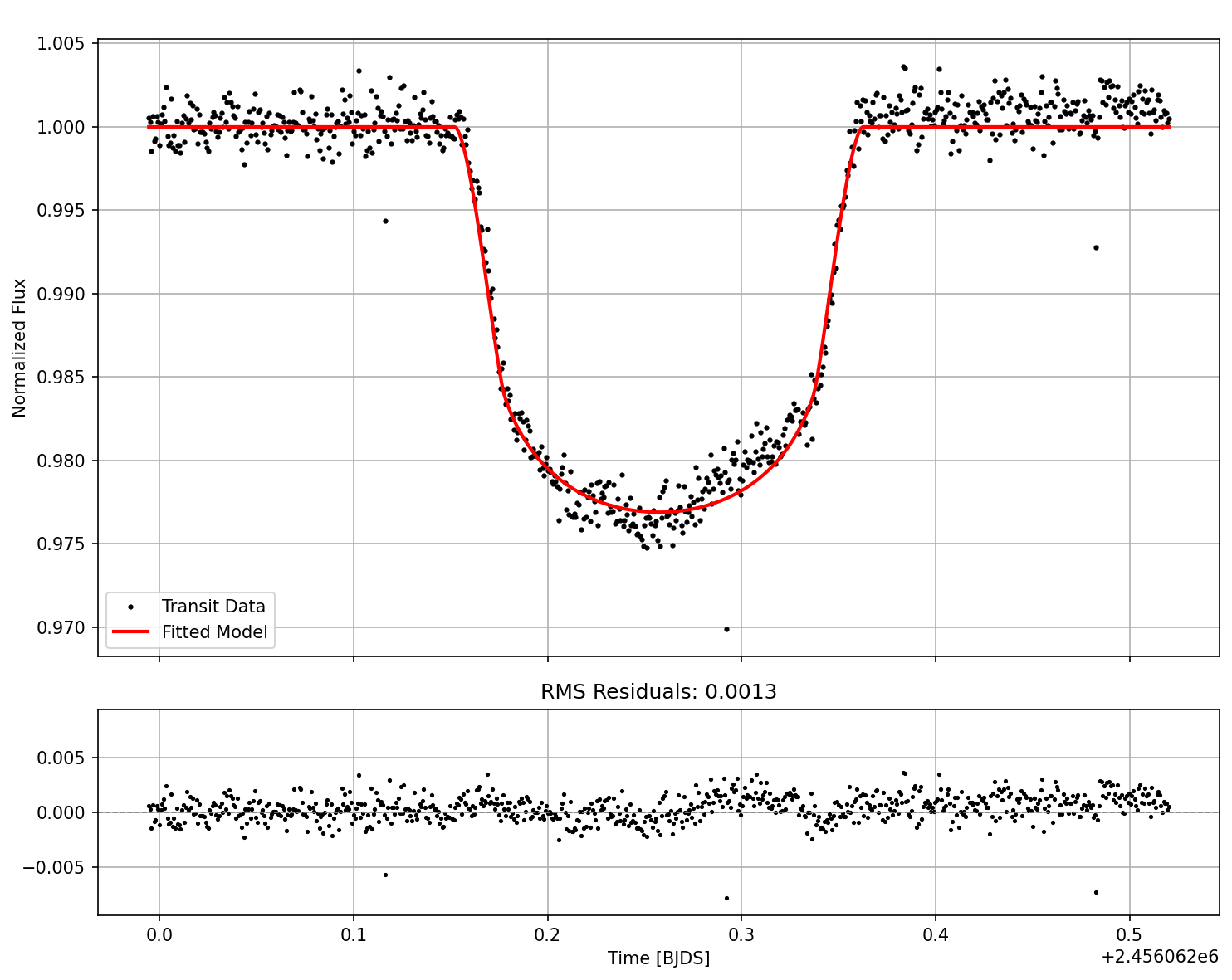}\\
Kepler-468 b (1229.303)
\end{minipage}\hfill
\begin{minipage}[t]{0.18\textwidth}
\centering
\includegraphics[width=\textwidth]{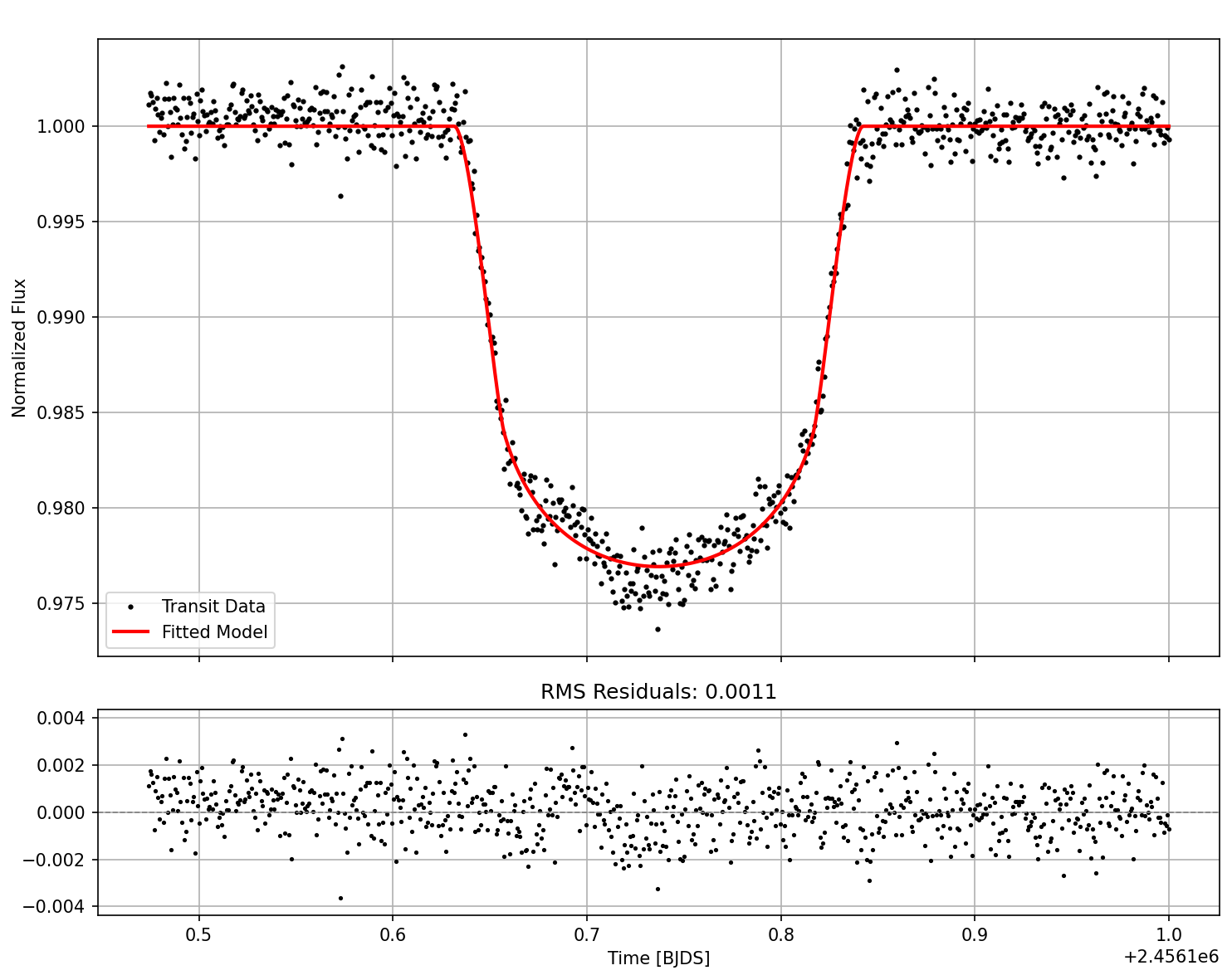}\\
Kepler-468 b (1267.693)
\end{minipage}

\vspace{10pt}

\begin{minipage}[t]{0.18\textwidth}
\centering
\includegraphics[width=\textwidth]{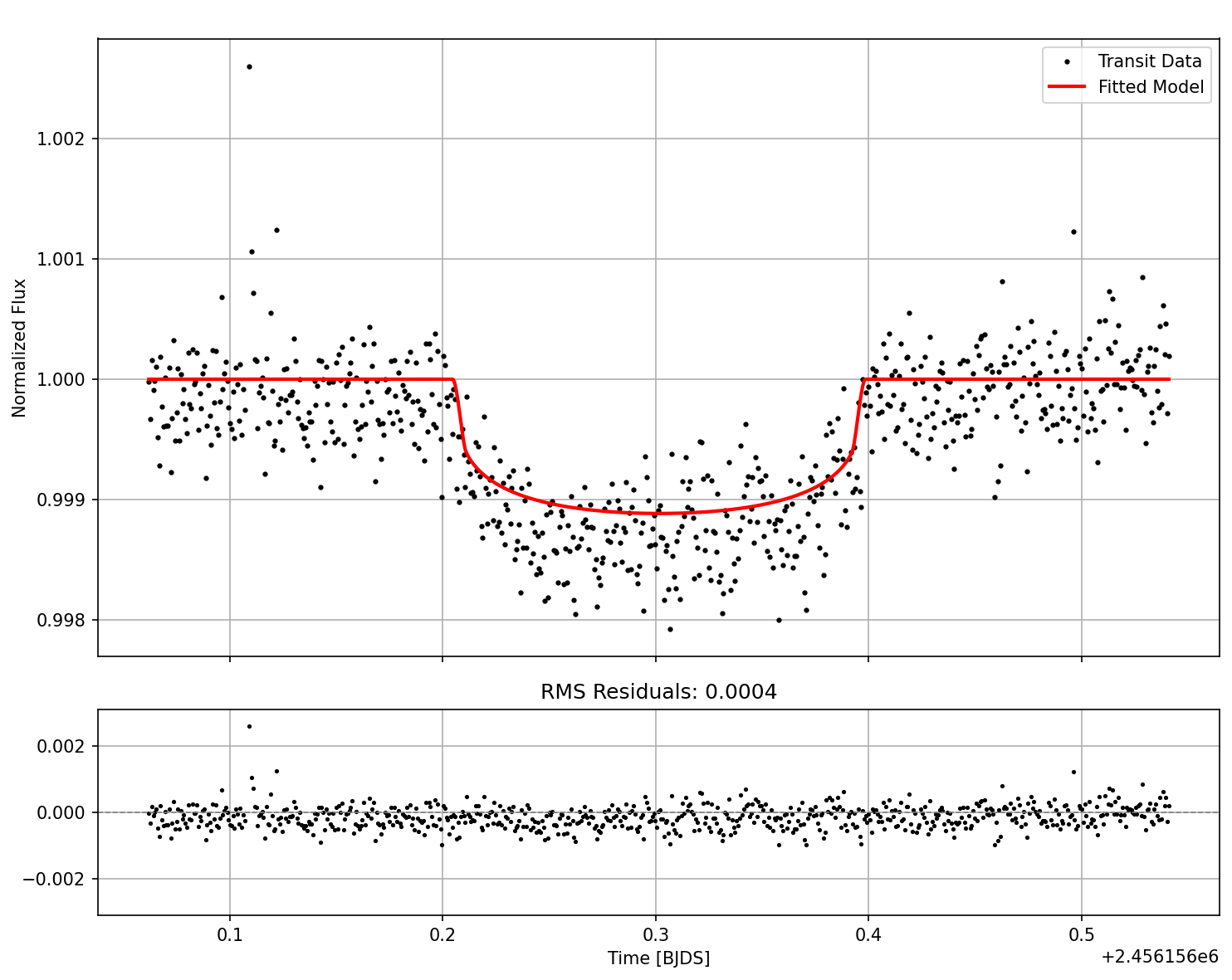}\\
Kepler-396 b (1323.333)
\end{minipage}\hfill
\begin{minipage}[t]{0.18\textwidth}
\centering
\includegraphics[width=\textwidth]{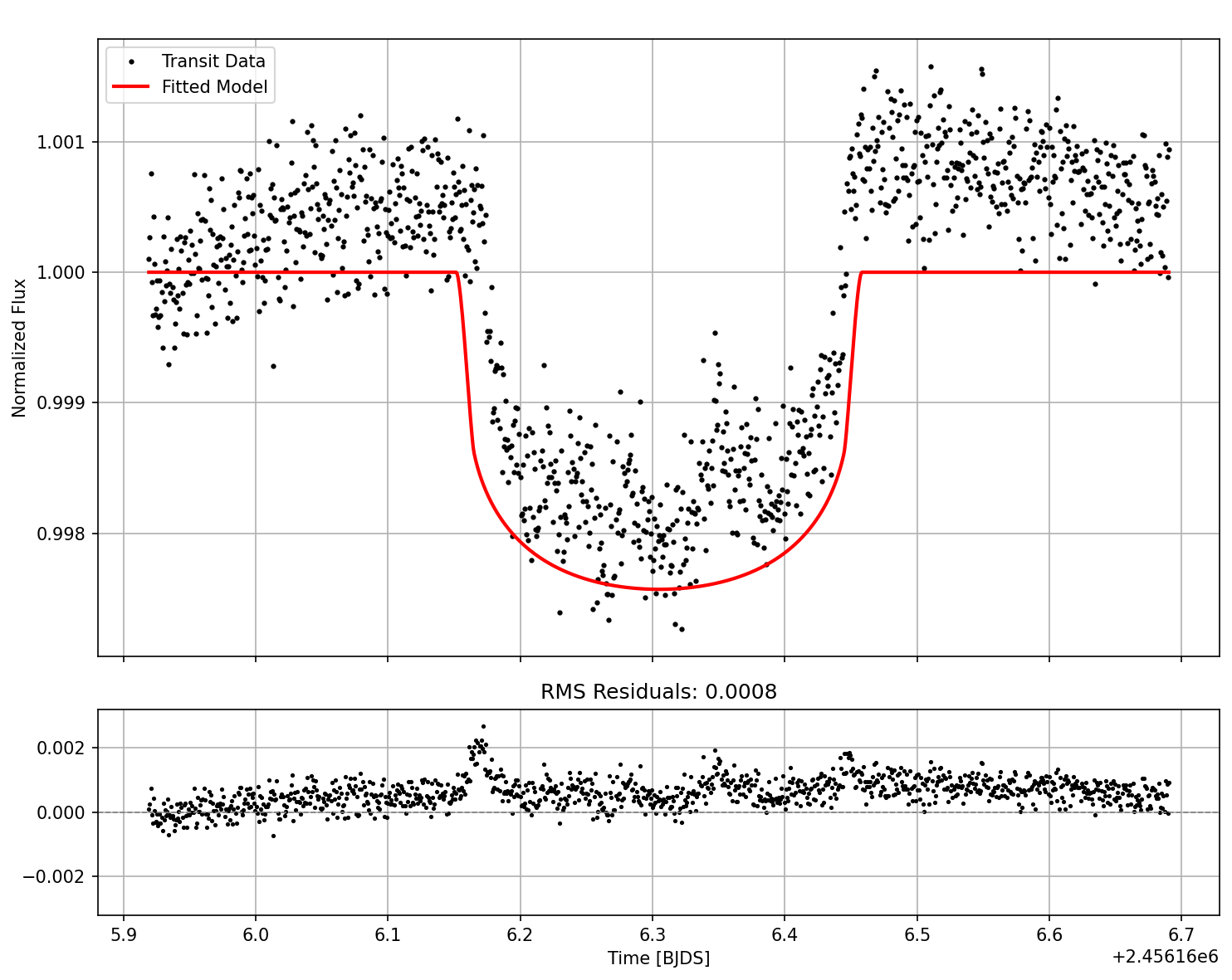}\\
Kepler-396 c (1333.349)
\end{minipage}\hfill
\begin{minipage}[t]{0.18\textwidth}
\centering
\includegraphics[width=\textwidth]{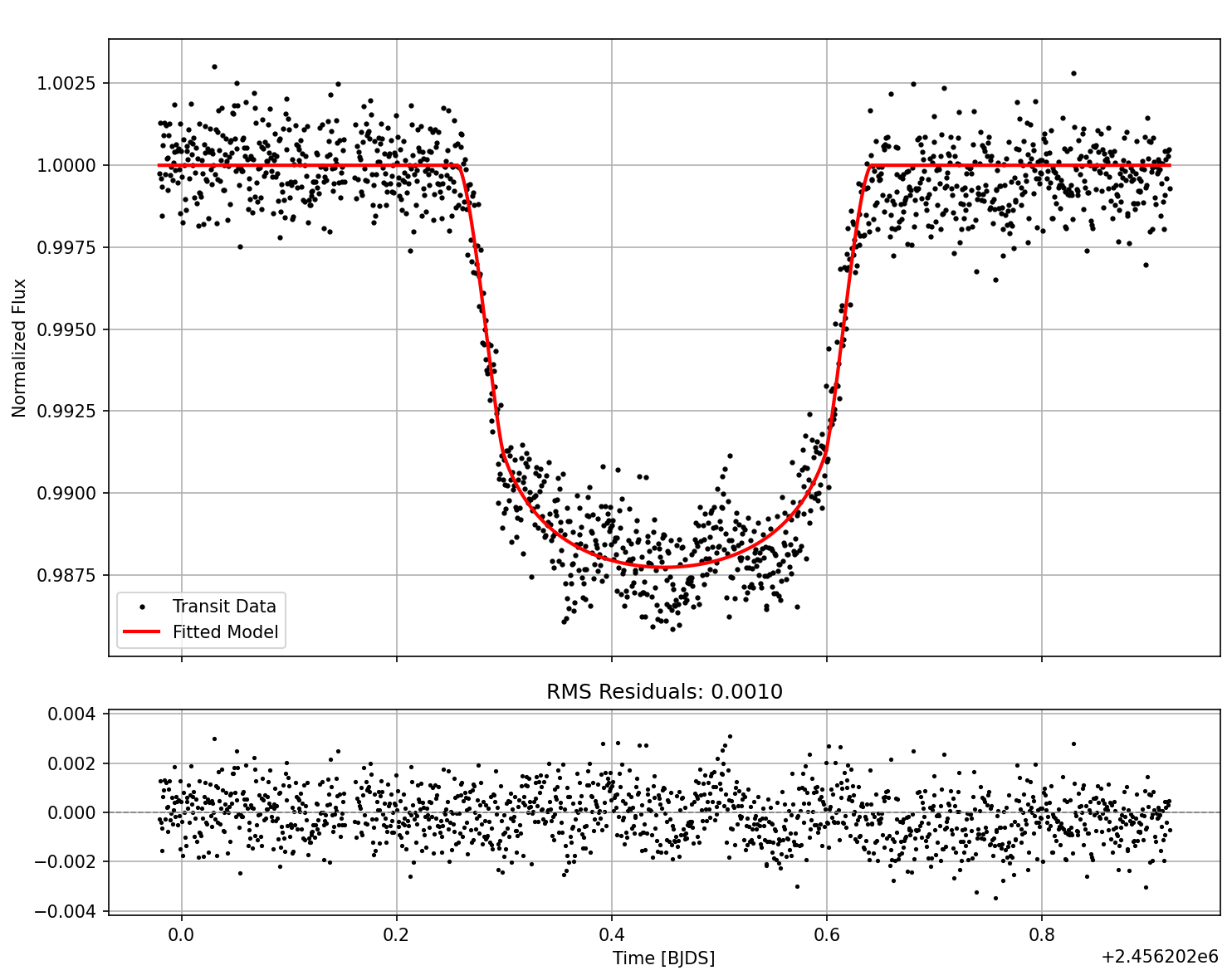}\\
Kepler-289 c (1369.499)
\end{minipage}\hfill
\begin{minipage}[t]{0.18\textwidth}
\centering
\includegraphics[width=\textwidth]{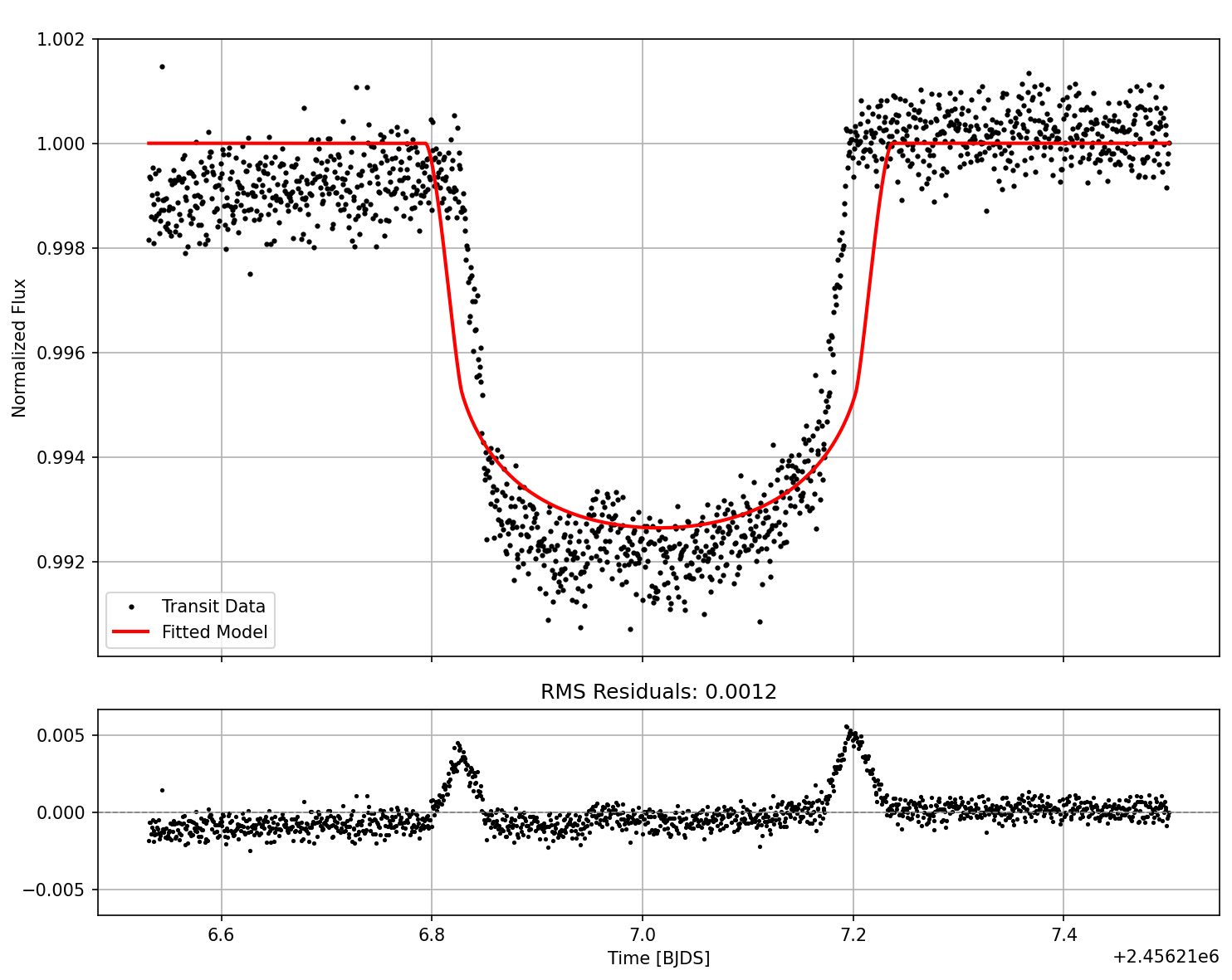}\\
Kepler-539 b (1383.969)
\end{minipage}\hfill
\begin{minipage}[t]{0.18\textwidth}
\centering
\includegraphics[width=\textwidth]{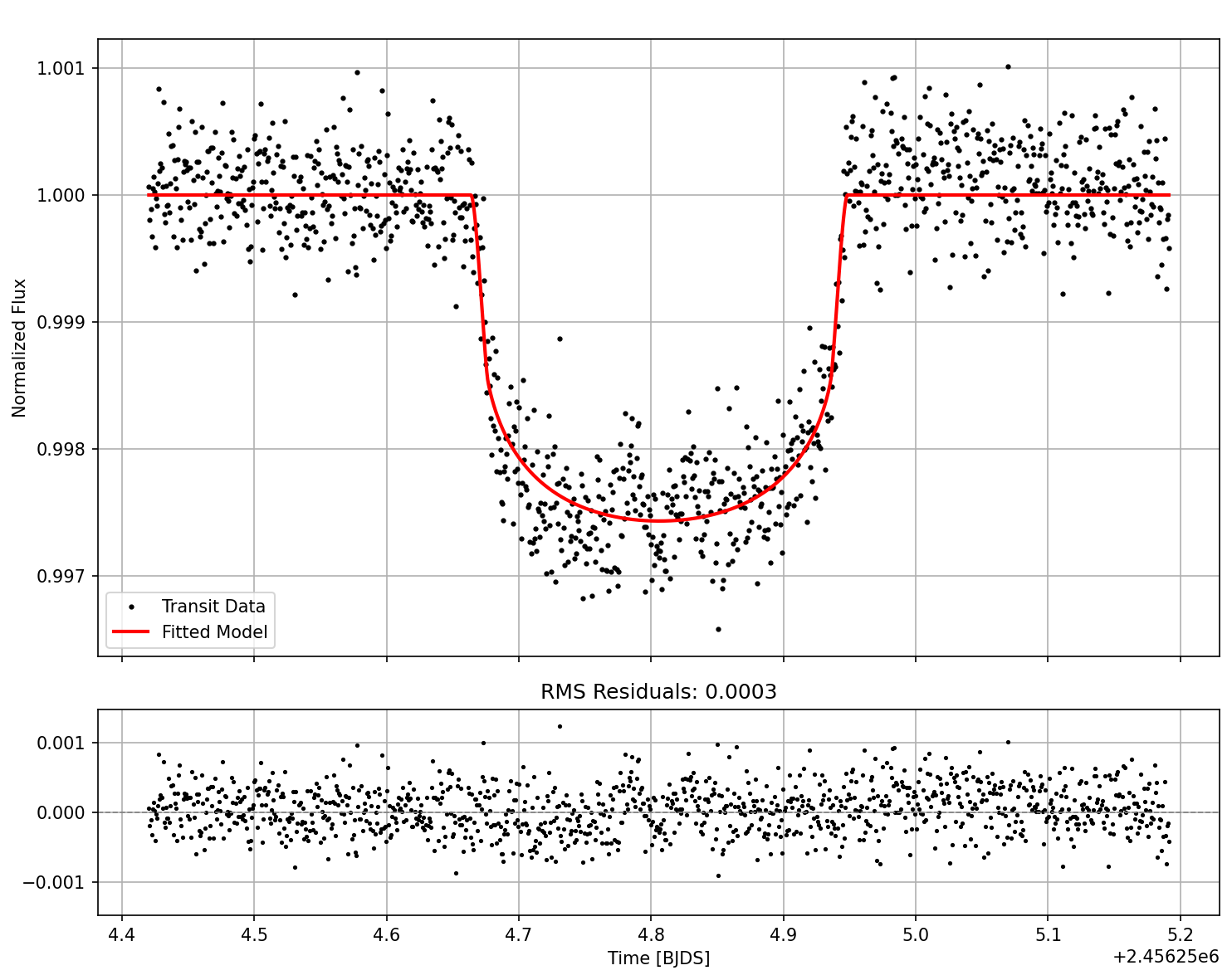}\\
Kepler-396 c (1421.786)
\end{minipage}

\vspace{10pt}

\begin{minipage}[t]{0.18\textwidth}
\centering
\includegraphics[width=\textwidth]{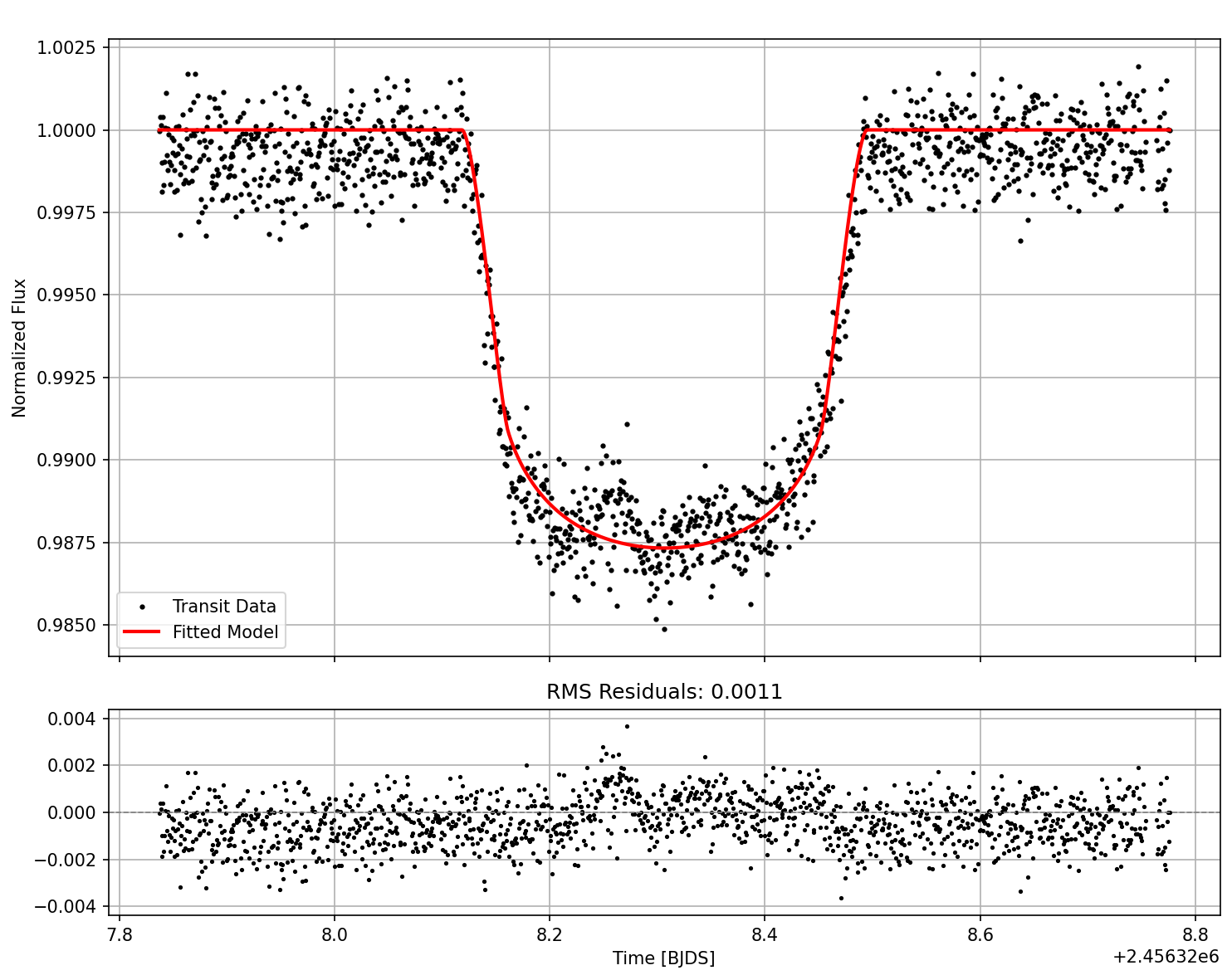}\\
Kepler-289 c (1495.264)
\end{minipage}\hfill
\begin{minipage}[t]{0.18\textwidth}
\centering
\includegraphics[width=\textwidth]{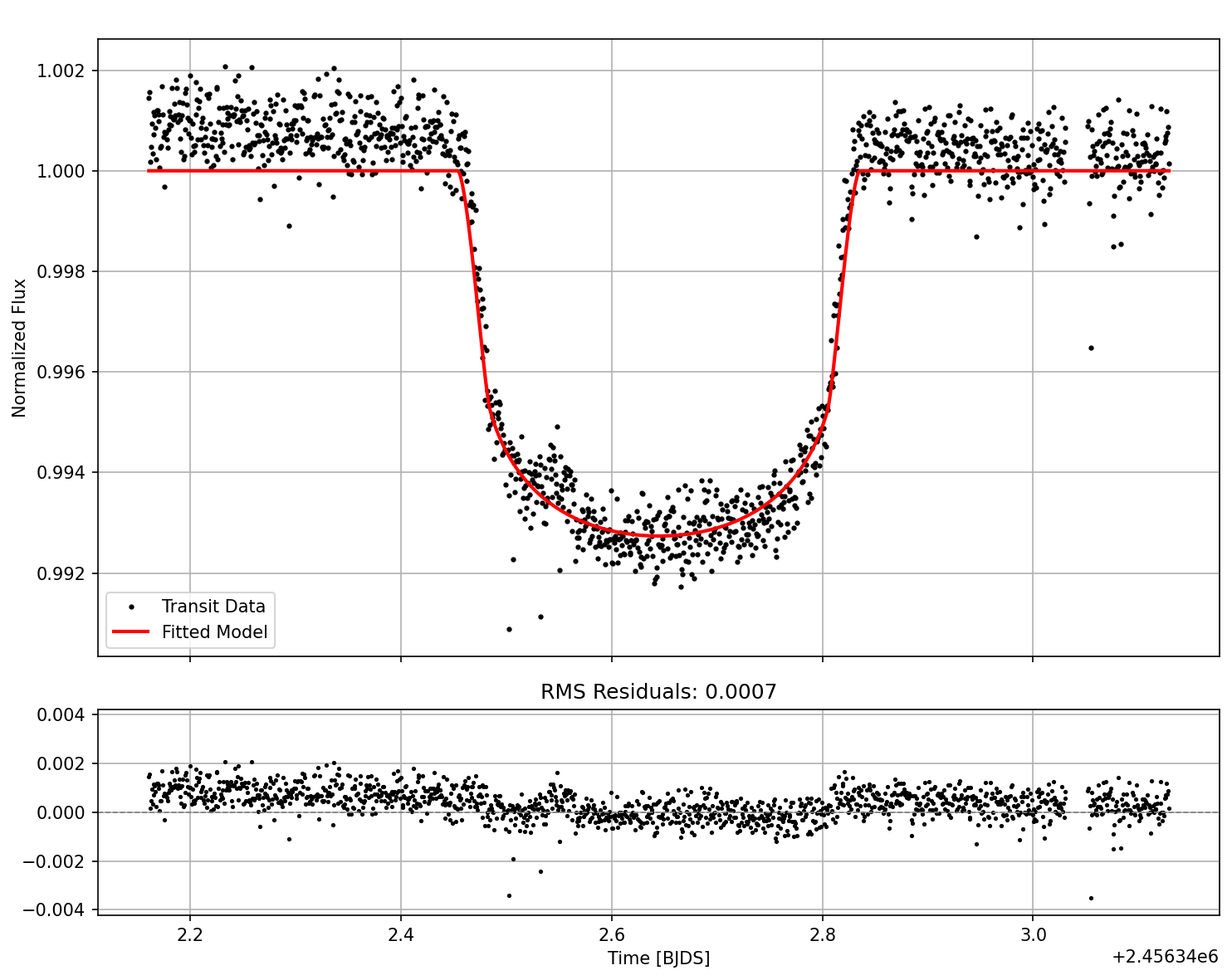}\\
Kepler-539 b (1509.551)
\end{minipage}\hfill
\begin{minipage}[t]{0.18\textwidth}
\centering
\includegraphics[width=\textwidth]{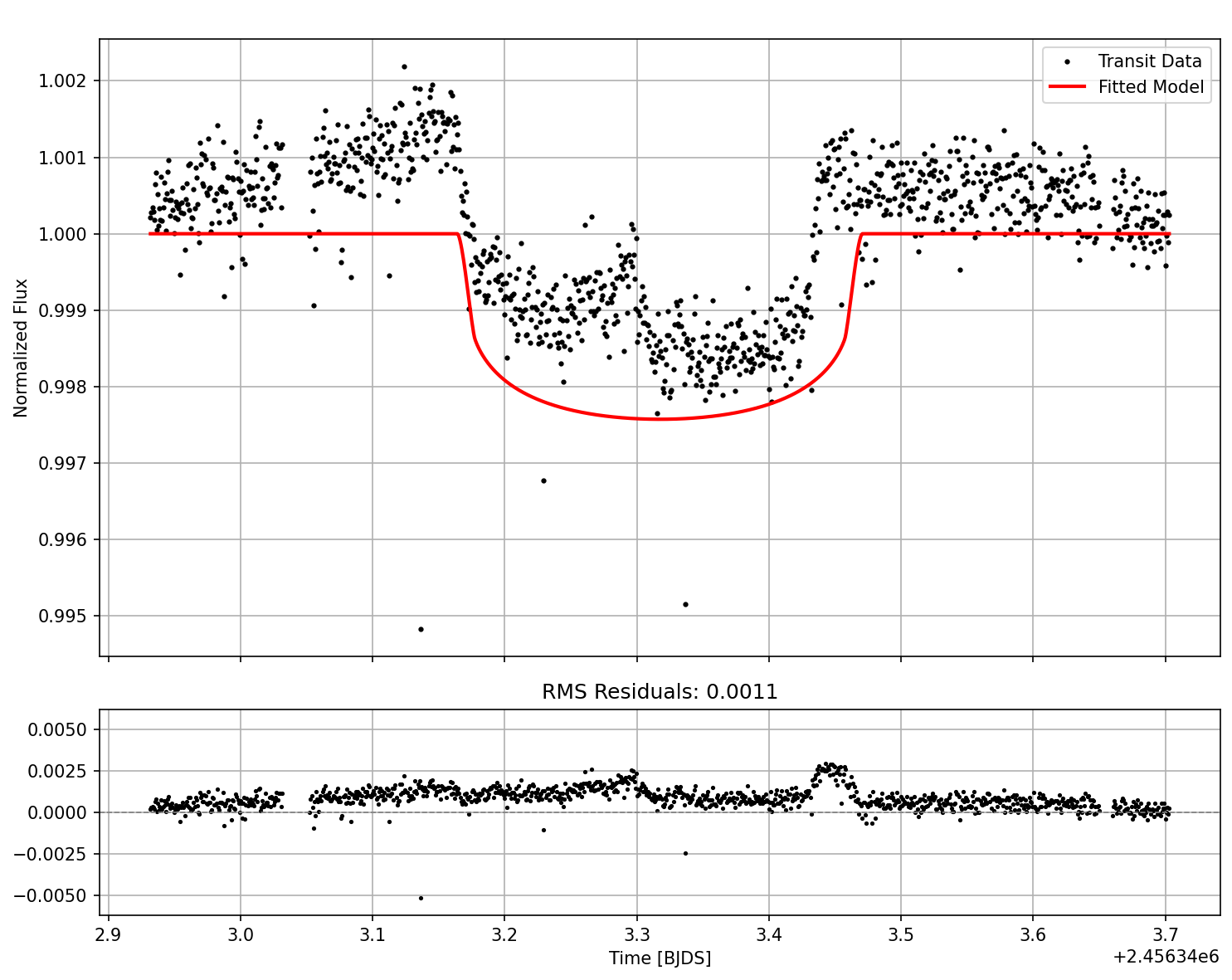}\\
Kepler-396 c (1510.285)
\end{minipage}\hfill
\begin{minipage}[t]{0.18\textwidth}
\centering
\includegraphics[width=\textwidth]{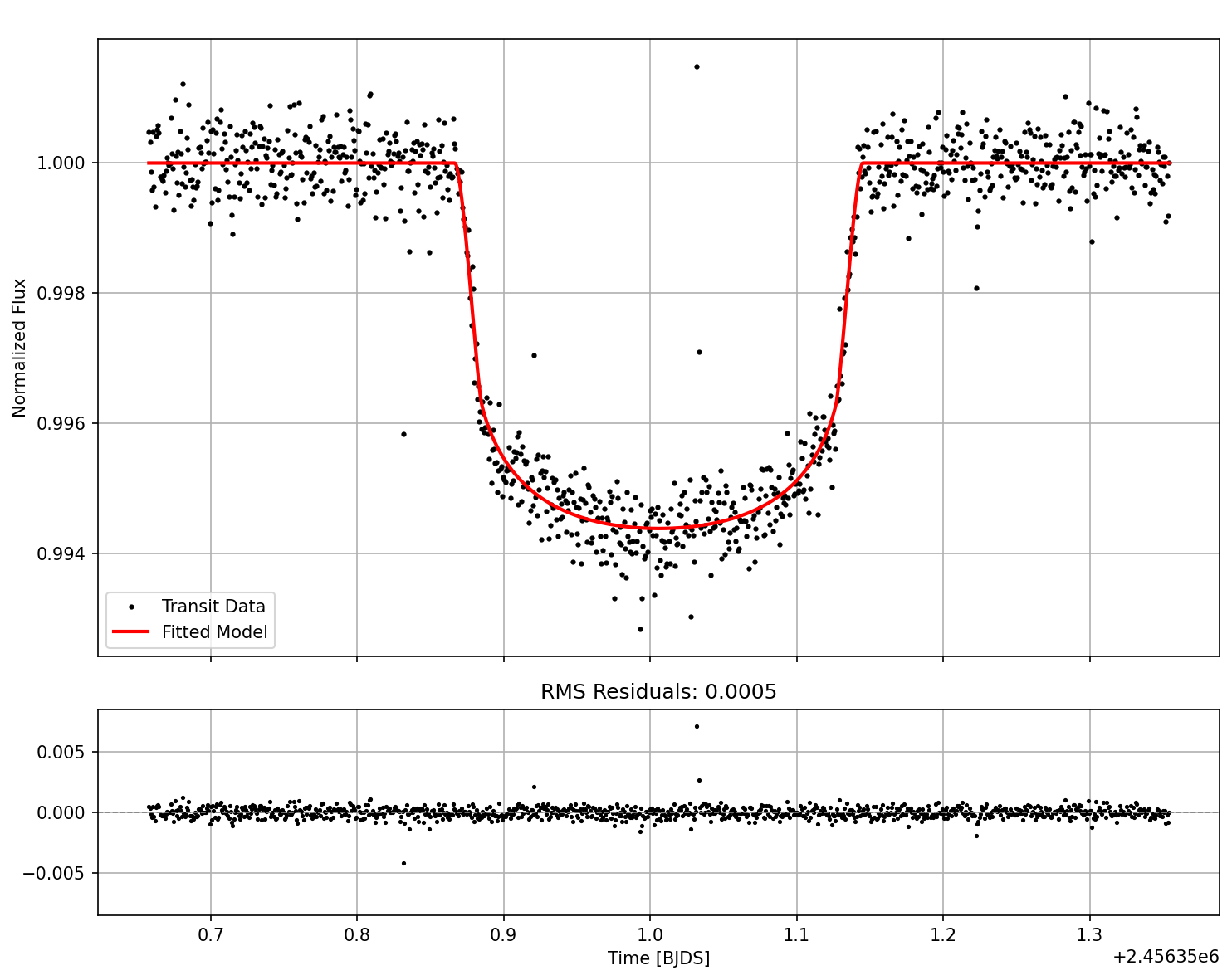}\\
Kepler-89 d (1518.032)
\end{minipage}\hfill
\begin{minipage}[t]{0.18\textwidth}
\centering
~\\
~
\end{minipage}

\caption{Detected SSCE-like signals. The central time of the event (BKJD) is indicated alongside the planet identifier.}
\label{fig:all_SSCEs}
\end{figure*}

We emphasize that while this statistical framework robustly confirms the presence of a signal above the correlated noise floor, it does not uniquely determine its physical origin. A low $p$-value guarantees that the flux anomaly is statistically significant, but it cannot independently distinguish a true SSCE from other transient phenomena, such as a low-energy flare or an isolated instrumental systematic, that might present a similar morphology. Therefore, the astrophysical classification of these validated signals as SSCEs relies fundamentally on the expert vetting phase. In Table~\ref{tab:SSCEs_detected} and Fig.~\ref{fig:all_SSCEs}, we present all SSCE-like signals detected in our light curve inspection.

While an ideal SSCE detection comprises a visually distinct anomaly, statistical significance ($p < 0.05$), and clear stellar rotational modulation, we also retained: (i) marginal signals ($p > 0.05$) in stars with confirmed rotation, provided the star exhibits other statistically validated SSCEs in different transits (Kepler-396 b at BKJD 1323.334, Kepler-468 b at BKJD 459.662, Kepler-539 b at BKJD 1383.969); (ii) isolated, statistically significant ($p < 0.05$) anomalies in stars lacking clear rotational modulation (Kepler-89 d at BKJD 1518.032); and (iii) signals with marginal significance ($p > 0.05$) and no rotational context, provided their visual aspect is compelling (Kepler-419 b at BKJD 893.353) or they appear across multiple transits  (KOI-00130.01 at BKJD 394.671, 497.227, and 531.418). This multi-tiered approach ensures that potentially genuine anomalies are not prematurely discarded.

\begin{table}[ht]
\centering
\small
\setlength{\tabcolsep}{3.5pt}
\caption{List of SSCE-like signals detected, along with their estimated center times ($t_{\rm c}$), $p$-values, and corresponding curve and transit IDs for the processed curves. The ($\dagger$) symbol labels events not found in previous studies.}
\label{tab:SSCEs_detected}
\begin{tabular}{lcccc}
\toprule
Exoplanet & $t_{\rm c}$ (BKJD) & $p$-value & Curve ID & Transit ID \\
\midrule
Kepler-89 d ($\dagger$)        & 1518.032 & 0.011    & 171 & 1 \\
Kepler-289 c ($\dagger$)       & 865.920  & 0.017    & 229 & 0 \\
Kepler-289 c ($\dagger$)       & 1369.499 & 0.043    & 3   & 0 \\
Kepler-289 c ($\dagger$)       & 1495.264 & 0.036    & 92  & 0 \\
Kepler-396 b ($\dagger$)       & 1323.334 & 0.145    & 69  & 0 \\
Kepler-396 c ($\dagger$)       & 1156.298 & $<0.001$ & 231 & 0 \\
Kepler-396 c ($\dagger$)       & 1333.349 & 0.033    & 93  & 0 \\
Kepler-396 c ($\dagger$)       & 1421.786 & 0.037    & 14  & 0 \\
Kepler-396 c ($\dagger$)       & 1510.286 & $<0.001$ & 232 & 0 \\
Kepler-411 c ($\dagger$)       & 1044.013 & $<0.001$ & 112 & 5 \\
Kepler-411 c (\#1)             & 1051.825 & $<0.001$ & 197 & 6 \\
Kepler-411 c (\#2)             & 1051.857 & 0.006    & 197 & 6 \\
Kepler-411 c                   & 1083.167 & 0.013    & 197 & 10 \\
Kepler-419 b ($\dagger$)       & 893.353  & 0.239    & 221 & 1 \\
Kepler-468 b ($\dagger$)       & 305.757  & 0.033    & 124 & 1 \\
Kepler-468 b (\#1, $\dagger$)  & 459.662  & 0.270    & 124 & 5 \\
Kepler-468 b (\#2, $\dagger$)  & 459.706  & 0.049    & 124 & 5 \\
Kepler-468 b ($\dagger$)       & 498.142  & 0.022    & 124 & 6 \\
Kepler-468 b ($\dagger$)       & 1190.816 & 0.032    & 202 & 0 \\
Kepler-468 b ($\dagger$)       & 1229.303 & 0.040    & 202 & 1 \\
Kepler-468 b ($\dagger$)       & 1267.694 & 0.048    & 202 & 2 \\
Kepler-539 b ($\dagger$)       & 1383.969 & 0.159    & 201 & 0 \\
Kepler-539 b ($\dagger$)       & 1509.552 & 0.031    & 59  & 0 \\
KOI00130.01 ($\dagger$)        & 394.671  & 0.145    & 147 & 3 \\
KOI00130.01 ($\dagger$)        & 497.227  & 0.068    & 147 & 6 \\
KOI00130.01 ($\dagger$)        & 531.418  & 0.161    & 147 & 7 \\
\bottomrule
\end{tabular}
\end{table}

Kepler-468 b was initially cataloged as an eclipsing binary  \citet{2011AJ....142..160S} and subsequently validated \citep{2016ApJ...822...86M}. Six transits of this system show SSCE-like events with a $p$-value $< 0.05$. The light curve exhibits obvious rotational modulation; however, we found no determinations in the literature for the rotation period of the host star, which we estimated to be $P_{\rm rot}=11.24 \pm 0.12$ d, with false alarm probability (FAP) $\ll$ 1, using a Generalized Lomb-Scargle (GLS) analysis \citep{2009A&A...496..577Z}.

KOI-130.01 is listed as a false positive in the NEA, although \citet{2016ApJ...822...86M} assigned it candidate status with FAP of 4.1\%. The light curve does not exhibit rotational modulation. We found no determinations for the stellar rotation period in the literature, and our GLS analysis did not yield any significant period. Nevertheless, we report three marginal flux enhancements ($p$-value $> 0.05$) observed during different transits of this object.

We detected three signals consistent with SSCEs ($p$-value $< 0.05$) during three transits of Kepler-289 c. The host star, Kepler-289, is known to be magnetically active and exhibits clear rotational modulation; \citet{2023A&A...678A..24R} derived a rotation period of $P_{\mathrm{rot}} = 8.75 \pm 0.84$ d. Furthermore, the occurrence of stellar flares in this system has been documented by \citet{2022ApJS..261...26S}.

Kepler-396 is another active star exhibiting clear rotational modulation, with a period of $P_{\mathrm{rot}} = 13.19 \pm 0.84$ d \citep{2023A&A...678A..24R} and flaring activity. We detected four signals consistent with SSCEs during four transits of Kepler-396 c (all with a $p$-value $< 0.05$), as well as an additional marginal event during a transit of Kepler-396 b.

Kepler-411 is an active star that has been extensively studied, with $P_{\mathrm{rot}} = 10.44 \pm 0.71$ d \citep{2023A&A...678A..24R}. Based on both short- and long-cadence \textit{Kepler} light curves, \citet{2021ApJ...922L..23A} reported a total of 198 SSCEs to model its starspot structure, although they did not explicitly publish the event timings. Subsequently, \citet{2026A&A...708A...8T} also investigated the spot architecture using the SSCEs reported by \citet{2021ApJ...922L..23A}. 
In this work, we detected four events consistent with SSCEs ($p$-value $< 0.05$), three of which (two in transit at BKJD~5884.82 and one at 5916.16) were previously reported by \citet{2025A&A...704A.102H}.

Kepler-539 exhibits clear rotational modulation with a period of $P_{\mathrm{rot}} = 11.87 \pm 0.78$ d \citep{2023A&A...678A..24R}. In this work, we identified two SSCE-like events during transits of Kepler-539 b. Notably, \citet{2016A&A...590A.112M} had already suggested the potential contamination of this planet's transits by starspots in the context of their study on TTVs.

\subsubsection{Flares}

We detected a total of seven stellar flares occurring shortly before, during, or immediately after the transits of three exoplanets: Kepler-396 b (one event), Kepler-411 b (five events), and Kepler-411 c (one event).

None of these events are reported in the catalogs by \citet{2017ApJS..232...26V} and \citet{2019ApJS..241...29Y}, which rely on long-cadence light curves, nor do they appear in the systematic searches by \citet{2016ApJ...829...23D} and \citet{2022ApJS..261...26S}, which incorporate both long- and short-cadence data. However, four of these events (at BKJD 1077.718, 1468.390, 1564.552, and 1579.579) are tabulated in the study by \citet{2021ApJ...922L..23A} regarding the magnetically active star Kepler-411. To the best of our knowledge, the remaining three events are undocumented in the existing literature.

\begin{figure*}[ht]
\centering
\scriptsize

\begin{minipage}[b]{0.24\textwidth}
\centering
\includegraphics[width=\textwidth]{Figures/Flares/Kepler411_b_2455910_718.pdf}\\
Kepler-411 b (1077.718)
\end{minipage}
\hfill
\begin{minipage}[b]{0.24\textwidth}
\centering
\includegraphics[width=\textwidth]{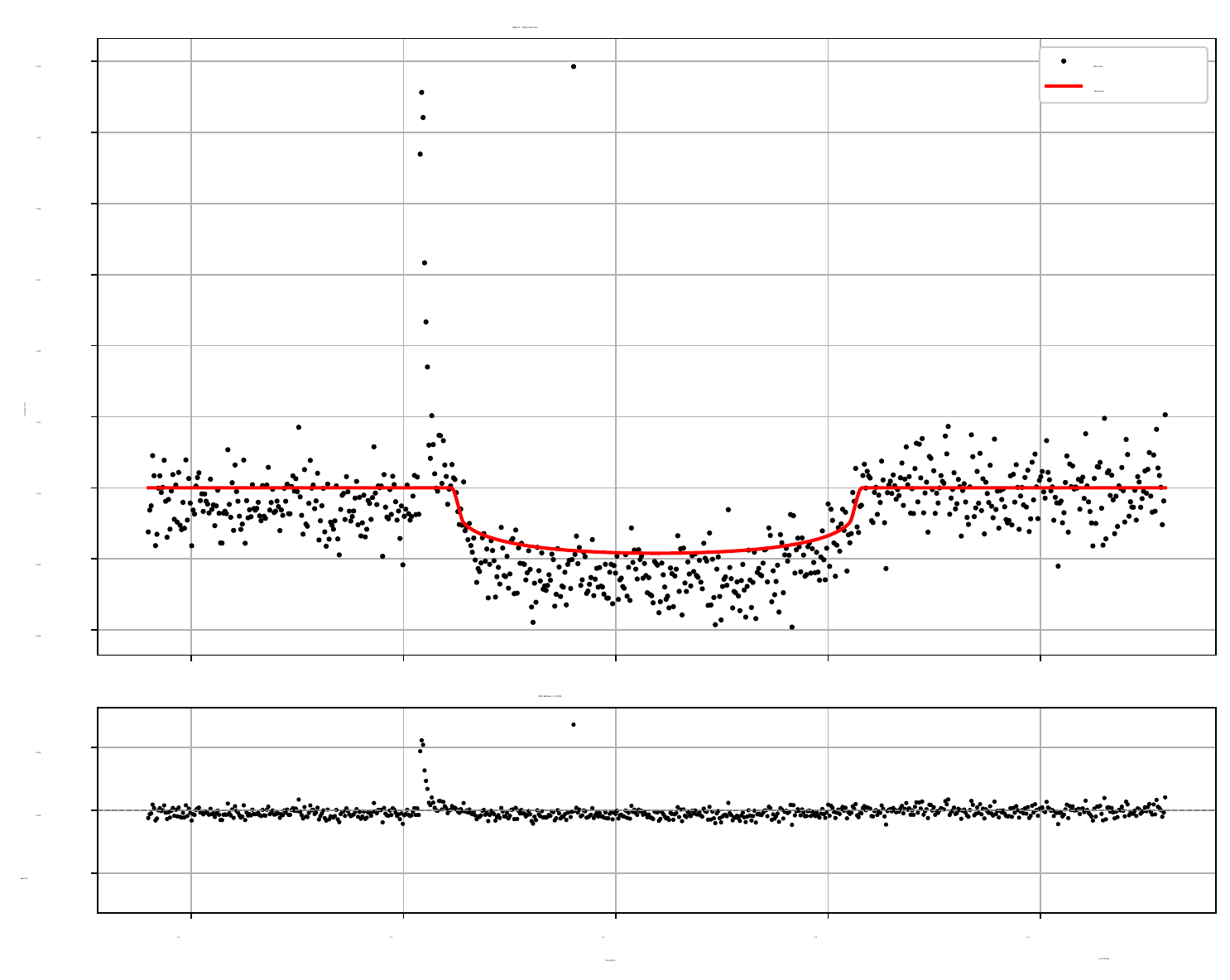}\\
Kepler-396 b (1108.319)
\end{minipage}
\hfill
\begin{minipage}[b]{0.24\textwidth}
\centering
\includegraphics[width=\textwidth]{Figures/Flares/Kepler411_b_2456184_207.pdf}\\
Kepler-411 b (1351.207)
\end{minipage}
\hfill
\begin{minipage}[b]{0.24\textwidth}
\centering
\includegraphics[width=\textwidth]{Figures/Flares/Kepler411_c_2456253_058.pdf}\\
Kepler-411 c (1420.058)
\end{minipage}

\vspace{10pt}

\begin{minipage}[b]{0.24\textwidth}
\centering
\includegraphics[width=\textwidth]{Figures/Flares/Kepler411_b_2456301_39.pdf}\\
Kepler-411 b (1468.390)
\end{minipage}
\hspace{8pt}
\begin{minipage}[b]{0.24\textwidth}
\centering
\includegraphics[width=\textwidth]{Figures/Flares/Kepler411_b_2456397_552.pdf}\\
Kepler-411 b (1564.552)
\end{minipage}
\hspace{8pt}
\begin{minipage}[b]{0.24\textwidth}
\centering
\includegraphics[width=\textwidth]{Figures/Flares/Kepler411_b_2456412_579.pdf}\\
Kepler-411 b (1579.579)
\end{minipage}

\caption{Flare events detected in this work. The central time of the transit (BKJD) is indicated alongside the planet identifier.}
\label{fig:all_flares}
\end{figure*}

These flares are listed in Table~\ref{tab:flares_detected}, and the transits in which they occur are shown in Fig.~\ref{fig:all_flares}.

\begin{table}[ht]
\centering
\small
\setlength{\tabcolsep}{6pt}
\caption{List of transits with stellar flares occurring shortly before, during, or immediately after the transit detected in this work.}
\label{tab:flares_detected}
\begin{tabular}{lccc}
\toprule
Exoplanet & $t_{\rm c}$ (BKJD) & Curve ID & Transit ID \\
\midrule
Kepler-396 b  & 1108.319 & 91  & 0 \\
Kepler-411 b  & 1077.718 & 142 & 25 \\
Kepler-411 b  & 1351.207 & 169 & 18 \\
Kepler-411 b  & 1468.390 & 157 & 16 \\
Kepler-411 b  & 1564.552 & 148 & 1 \\
Kepler-411 b  & 1579.579 & 148 & 6 \\
Kepler-411 c  & 1420.058 & 198 & 0 \\
\bottomrule
\end{tabular}
\end{table}

\subsubsection{Transit Duration Variations}

We identified 38 transits exhibiting transit duration variation (TDV) anomalies, indicating discrepancies between the observed and predicted durations. We classified these cases into two categories.

The first category comprises 31 transits belonging to systems that exhibit systematic duration deviations across all observed light curves. These systems are characterized by minimal stellar variability---which rules out artifacts introduced during the detrending process---along with low or zero orbital eccentricities, and architectures lacking strong dynamical interactions (i.e., single-planet or dynamically quiet multi-planet systems). In these instances, the systematic nature of the deviations suggests that the baseline parameters, specifically the fixed orbital inclinations adopted from the NEA, are suboptimal. To isolate and correct this effect for the affected systems (Kepler-108 b, Kepler-450 b, Kepler-458 c, and Kepler-533 b), we performed a secondary analysis. First, we derived a more accurate central time ($t_{\mathrm{c}}$) for each transit by applying a five-point moving average within a 12-hour window centered on the expected $t_{\mathrm{c}}$. Subsequently, we treated the orbital inclination as a free parameter and refitted the scaled semi-major axis ($a/R_{\star}$) and the planet-to-star radius ratio ($R_{\mathrm{p}}/R_{\star}$) to identify the configuration that minimized the TDVs.

In Table~\ref{tab:ajuste_parametros} we report the results of this refit, yielding lower $\chi^2$ values and significantly reducing the TDV amplitudes across all evaluated events. Consequently, we suggest that the nominal NEA parameters for these four systems require revision. Specifically, our fits favor a lower orbital inclination for Kepler-108 b and Kepler-458 c (averaging $89.2^\circ$ and $85.5^\circ$, respectively), a slightly higher inclination for Kepler-450 b ($89.3^\circ$), and a reduced scaled semi-major axis for Kepler-533 b ($a/R_{\star} = 34.00$).

The second category encompasses the remaining seven transits, associated with Kepler-411 c and Kepler-419 b (four and three events respectively). Both targets reside in complex multi-planetary architectures with well-documented TTVs and significant stellar variability. Properly disentangling genuine TDVs from the underlying dynamical interactions and stellar activity would require comprehensive photodynamical modeling. Consequently, a detailed characterization of these specific anomalies is beyond the scope of the present work.

\begin{table}[!htbp]
\centering
\scriptsize
\setlength{\tabcolsep}{3.5pt}
\caption{Best-fit parameters for systems exhibiting systematic TDVs. The rows corresponding to the planet names show the nominal parameters from the NEA. The label $i_{c}$ denotes fits with the inclination fixed to the NEA value, whereas $i_{f}$ indicates fits with the inclination left as a free parameter.}
\label{tab:ajuste_parametros}
\begin{tabular}{llccccc}
\toprule
Planet / Curve \# & Fit & $i$ ($^\circ$) & $a/R_{\star}$ & $R_{\rm p}/R_{\star}$ & TDV (min) & $\chi^2$ \\
\midrule
Kepler-108b & --           & 89.94 & 28.73 & 0.0375 & --     & -- \\
\# 81       & $i_{\rm c}$ & 89.94 & 28.73 & 0.0333 & 101.2  & 0.00059 \\
            & $i_{\rm f}$ & 89.56 & 32.40 & 0.0354 & -3.8   & 0.00050 \\
\# 139      & $i_{\rm c}$ & 89.94 & 28.73 & 0.0331 & 100.2  & 0.00322 \\
            & $i_{\rm f}$ & 88.94 & 28.73 & 0.0362 & 2.1    & 0.00278 \\
\# 178      & $i_{\rm c}$ & 89.94 & 28.73 & 0.0328 & 101.1  & 0.00109 \\
            & $i_{\rm f}$ & 89.19 & 30.87 & 0.0358 & -13.5  & 0.00088 \\
\midrule
Kepler-450b & --           & 88.82 & 26.38 & 0.0346 & --     & -- \\
\# 38       & $i_{\rm c}$ & 88.82 & 26.38 & 0.0350 & -60.6  & 0.00131 \\
            & $i_{\rm f}$ & 89.32 & 26.38 & 0.0335 & -8.6   & 0.00111 \\
\# 123      & $i_{\rm c}$ & 88.82 & 26.38 & 0.0374 & -58.6  & 0.00048 \\
            & $i_{\rm f}$ & 89.32 & 26.38 & 0.0359 & -7.6   & 0.00038 \\
\# 135      & $i_{\rm c}$ & 88.82 & 26.38 & 0.0354 & -61.6  & 0.00032 \\
            & $i_{\rm f}$ & 89.32 & 26.38 & 0.0343 & -8.6   & 0.00027 \\
\# 138      & $i_{\rm c}$ & 88.82 & 26.38 & 0.0354 & -60.6  & 0.00062 \\
            & $i_{\rm f}$ & 89.25 & 26.38 & 0.0341 & -13.5  & 0.00052 \\
\midrule
Kepler-458c & --           & 88.27 & 10.28 & 0.0299 & --     & -- \\
\# 42       & $i_{\rm c}$ & 88.27 & 10.31 & 0.0254 & 261.5  & 0.02385 \\
            & $i_{\rm f}$ & 85.58 & 10.28 & 0.0254 & -44.4  & 0.02098 \\
\# 222      & $i_{\rm c}$ & 88.27 & 10.32 & 0.0254 & 261.5  & 0.00578 \\
            & $i_{\rm f}$ & 85.52 & 10.28 & 0.0254 & -57.2  & 0.00507 \\
\midrule
Kepler-533b & --           & 88.68 & 40.00 & 0.0431 & --     & -- \\
\# 11       & $i_{\rm c}$ & 88.68 & 40.00 & 0.0432 & -126.6 & 0.00146 \\
            & $i_{\rm f}$ & 88.68 & 34.00 & 0.0434 & -19.7  & 0.00121 \\
\# 37       & $i_{\rm c}$ & 88.68 & 40.00 & 0.0432 & -127.4 & 0.00460 \\
            & $i_{\rm f}$ & 88.68 & 34.00 & 0.0446 & -18.7  & 0.00370 \\
\bottomrule
\end{tabular}
\end{table}

\subsubsection{Transit Timing Variations}

Our fitting procedure identifies discrepancies between the predicted linear ephemerides and the observed transit mid-times. These deviations may arise either from inaccuracies in the cataloged ephemerides or from genuine dynamical interactions that induce physical advances or delays.

Although comprehensive TTV analysis is not among the main goals of this work---as these variations were not identified through visual inspection but computed automatically to match models with actual transits---we conducted a basic analysis of these results for planets with at least three transits. From the raw TTVs obtained for each planet, we subtracted a linear trend to correct for orbital period inaccuracies (a notable case being Kepler-1556 b). Once this trend was removed, we compared our results with the TTVs published by \citet{2016ApJS..225....9H}.

Out of the 51 planets analyzed in this study, 24 have TTVs reported. We successfully recovered 17 of these, while the remaining seven did not meet the minimum requirement of three transits in short cadence necessary for analysis. These results are summarized in Table~\ref{tab:comparativa_ttvs}. 
It is important to note that the discrepancies between the TTVs previously reported and this work are due to several factors: our study is limited to short cadence data, a shorter temporal baseline, a code that, while capable of handling TTVs for fitting purposes, is not specifically optimized for such analysis, and the simplified nature of our TTV estimation.

While obtaining maximum TTV values lower than literature values is expected---simply because our observational window may not capture the peak of the dynamical cycle---computing significantly higher values warrants a detailed evaluation. We note that in some instances, our higher measurements are supported by other studies. For example, our maximum TTV of 67 minutes for Kepler-419 b exceeds the 39 minutes reported by \citet{2016ApJS..225....9H}, but remains conservative when compared to the variations of up to three hours identified by \citet{2018A&A...615A..90A}. Similarly, our computed amplitude of 17 minutes for Kepler-79 d is in agreement with the 16-minute value derived by \citet{2018ApJS..234....9O}.

Conversely, for other targets, the overestimation highlights the limitations of utilizing a general-purpose fitting tool rather than a specialized photodynamical model. For Kepler-522 b, our maximum TTV (38 min) surpasses both, the 12-minute amplitude reported in \citet{2016ApJS..225....9H} catalog and the 20-minute amplitude reported by \citet{2025MNRAS.538.2283K}. The host star exhibits strong rotational modulation \citep[$P_{\mathrm{rot}} = 11.45 \pm 0.11$ d;][]{2021ApJS..255...17S}. We suggest that residual systematics from a suboptimal detrending process artificially shifted the transit mid-times during our fitting procedure. Methodological limitations also explain the discrepancies in Kepler-289 c and Kepler-11 e. For Kepler-289 c, the long transit duration ($\sim 0.38$ d) inherently broadens the minimum of the light curve, increasing the uncertainty in the exact mid-time determination and allowing the code to artificially inflate the TTV to absorb residual noise. A similar algorithmic compensation occurs in regimes of low signal-to-noise ratio, which accounts for the high 100-minute TTV value we obtained for Kepler-11 e.

\begin{table}[!htbp]
\centering
\small
\setlength{\tabcolsep}{3pt}
\caption{Comparison between maximum TTV values detected in this work and literature values from \citet{2016ApJS..225....9H}. For all planets with TTVs that were not recovered in this study, the available short cadence data contained fewer than the minimum requirement of three transits.}
\label{tab:comparativa_ttvs}
\begin{tabular}{llcc}
\toprule
KEPOI & \textit{Kepler} ID & $\rm TTV_{Holczer}$ (min) & $\rm TTV_{this\ work}$ (min) \\
\midrule
K01478.01 & Kepler-858 b  & 12.0  & 5.0   \\
K01474.01 & Kepler-419 b  & 39.0  & 67.0  \\
K00353.01 & ...            & 3.0   & ...    \\
K01781.03 & Kepler-411 d  & 40.0  & 42.0  \\
K02672.01 & Kepler-396 c  & 59.0  & 14.0  \\
K02672.02 & Kepler-396 b  & 32.0  & 6.0   \\
K00345.01 & Kepler-531 b  & 12.0  & 5.0   \\
K01783.01 & Kepler-1662 b & 6.0   & ...    \\
K00119.02 & Kepler-108 c  & 11.0  & ...    \\
K00319.01 & ...            & 13.0  & 16.0  \\
K01271.01 & ...            & 85.0  & ...    \\
K00139.01 & Kepler-111 c  & 211.0 & 28.0  \\
K00152.01 & Kepler-79 d   & 9.0   & 17.0  \\
K00318.01 & Kepler-522 b  & 12.0  & 38.0  \\
K01353.01 & Kepler-289 c  & 9.0   & 66.0  \\
K00075.01 & ...            & 20.0  & 20.0  \\
K00672.02 & Kepler-209 c  & 9.0   & 21.0  \\
K01242.01 & ...            & 3.0   & ...    \\
K00157.03 & Kepler-11 e   & 6.0   & 100.0 \\
K00372.01 & Kepler-539 b  & 14.0  & 3.0   \\
K00094.03 & Kepler-89 e   & 9.0   & 10.0  \\
K00638.01 & Kepler-199 b  & 100.0 & ...    \\
K00108.02 & Kepler-103 c  & 15.0  & 10.0  \\
K01335.01 & Kepler-820 b  & 6.0   & ...    \\
\bottomrule
\end{tabular}
\end{table}

\subsubsection{Concurrent Transits}

During the inspection process, an overlap was detected between the transits of Kepler-411 b and Kepler-411 c at BKJD=1059.678. Additionally, an overlap was identified between the transits of Kepler-89 e and Kepler-89 d at BKJD=378.521. Within the latter, an increase in flux is observed in the interior region, produced by a planet-planet eclipse, as discussed in \cite{2012ApJ...759L..36H}.

\section{Conclusions}

We presented the pilot phase of the DIPS-OjOS project, a pro-am collaboration designed for the systematic visual inspection of exoplanetary transits. The project aims to identify subtle photometric anomalies that may remain undetected or be misclassified by automated pipelines. To validate this approach, we developed a web-based interface that allowed trained collaborators to perform blind classifications of a sample of 917 transit events, including 362 simulated transits and 555 real short-cadence \textit{Kepler} transits. The high level of engagement resulted in 10\,645 individual classifications.

The simulated dataset, which included injected signals from starspots, satellites, and TTVs, enabled us to quantify the performance of the visual-inspection workflow. We established a baseline false-positive rate of 15.3\% and a recovery efficiency of 87.6\% for injected signals. These metrics allowed us to define a reliability scale for individual observers and to implement a weighted filtering strategy, reducing the number of real events requiring expert review.

The real sample consisted of short-cadence \textit{Kepler} data for 51 well-studied systems. From an initial subset of 363 transits flagged as suspicious by the collaborators, 88 were retained as viable candidates after expert validation. The remaining rejected cases were mainly associated with instrumental or astrophysical artifacts inherent to real archival data, highlighting the importance of combining visual inspection with expert vetting and simulated control samples.

Despite the deliberately limited scope of this pilot study, the inspection recovered a diverse set of candidate anomalies. We identified 26 SSCE-like signals across eight targets, 23 of which are previously unreported, and derived a rotation period for the host star of Kepler-468 b. We also identified seven stellar flares occurring in close temporal proximity to transits in two systems, including three events not found in previous catalogs. In addition, we reported 38 signals consistent with TDVs, of which 31 systematic deviations in four exoplanets suggest that their nominal orbital parameters may require refinement. Finally, we recovered a mutual transit event involving Kepler-89 d and e and obtained TTV measurements for 17 exoplanets, which we compared with previous specialized studies.

These results demonstrate that trained visual inspection, when combined with simulated controls, performance-based collaborator weighting, and expert validation, provides a feasible and scalable strategy for detecting subtle transit anomalies. The present work therefore establishes the operational and methodological basis for future phases of DIPS-OjOS, in which the workflow will be applied to larger samples from \textit{Kepler}, TESS, PLATO, and ground-based surveys with an expanded community of collaborators.

\section*{Data and code availability}

The web-based classification platform, project documentation, and the light-curve dataset used in this pilot phase are available at \url{https://dipsojos.uniovi.es}. The \textit{Kepler} photometry used in this work is publicly available from the NASA data archives. The custom analysis and simulation codes developed for this project are available from the authors upon reasonable request.

\begin{acknowledgments}
We thank Joaquín González-Nuevo and Sergio Luis Suárez Gómez for valuable comments on an earlier version of this manuscript. This work was supported by Agencia SEKUENS (Agencia de Ciencia, Competitividad Empresarial e Innovación Asturiana) through the grant for research groups of the Principality of Asturias (Ref. SEK-25-GRU-GIC-24-029); Plan Nacional of the Ministerio de Ciencia, Innovación y Universidades (Spain), grant number MCIU-22-PID2021-127331NB-I00; and HUNOSA through the Specific Collaboration Agreement for the Promotion of Research on Space Mining and Energy Resources, reference SV-21-HUNOSA-2. 
\end{acknowledgments}

% ...
%
\facilities{Kepler}

\software{
Python,
astropy \citep{2013A&A...558A..33A,2018AJ....156..123A,2022ApJ...935..167A},
astroquery \citep{2019AJ....157...98G},
lightkurve \citep{2018ascl.soft12013L},
wotan \citep{2019AJ....158..143H},
batman \citep{2015ascl.soft10002K}
}

\bibliography{references}
\bibliographystyle{aasjournalv7}
\end{document}